\pdfoutput=1
\documentclass[12pt,letterpaper]{article}

\usepackage[margin=1in]{geometry}
\usepackage[T1]{fontenc}
\usepackage[utf8]{inputenc}
\usepackage{mathptmx}

\usepackage{setspace}
\usepackage{ragged2e}
\usepackage{indentfirst}
\usepackage{graphicx}
\usepackage{booktabs}
\usepackage{array}
\usepackage{caption}
\usepackage{float}
\usepackage{pdflscape}
\usepackage{enumitem}

\usepackage{amsmath}
\usepackage{xcolor}
\usepackage[most]{tcolorbox}
\usepackage{listings}
\usepackage{multirow}
\usepackage{tabularx}
\usepackage{makecell}
\usepackage{setspace}
\usepackage{placeins}
\usepackage{longtable}
\usepackage{needspace}
\renewcommand{\ttfamily}{\rmfamily}
\usepackage[htt]{hyphenat}  %

\newcommand{\up}{}\newcommand{\down}{}        %
\newcolumntype{L}[1]{>{\raggedright\arraybackslash}p{#1}}
\newcolumntype{C}[1]{>{\centering\arraybackslash}p{#1}}
\AtBeginEnvironment{tcolorbox}{\par\singlespacing}
\makeatletter
\patchcmd{\@footnotetext}{\footnotesize}{\normalsize\singlespacing}{}{\errmessage{Footnote formatting patch failed}}
\makeatother
\usepackage[authoryear,round,longnamesfirst,sort]{natbib}
\setcitestyle{authoryear,round,semicolon,aysep={},yysep={,}}
\AtBeginEnvironment{thebibliography}{\doublespacing}

\usepackage[hidelinks,hypertexnames=false]{hyperref}
\RaggedRight
\renewcommand{\tablename}{TABLE}

\newcommand{\primaryheading}[1]{%
  \par\addvspace{\baselineskip}%
  {\centering\bfseries\MakeUppercase{#1}\par}%
  \nobreak\vspace{\baselineskip}%
}

\newcommand{\secondaryheading}[1]{%
  \par\vspace{\baselineskip}\noindent #1\par\vspace{\baselineskip}%
}

\newcommand{\tertiaryheading}[1]{%
  \par\vspace{\baselineskip}\noindent\hspace*{0.5in}\textit{#1.}\ %
}

\DeclareCaptionFont{samplecaption}{\rmfamily\normalsize\singlespacing}
\newcommand{\fignote}[1]{\par\vspace{6pt}{\singlespacing\RaggedRight\small\noindent NOTE.--- #1\par}}
\renewcommand{\tablename}{Table}

\begin{document}
\shortcites{Christiano2017,Jang2023,Ouyang2022,Shi2023,achiam2023gpt4,boussioux2024crowdless,brown2020language,bubeck2023sparks,cer2018universal,dellacqua2023jagged,galinsky2008power,ge2024scaling,gray2019forward,hao2026artificial,jackson2025ai,jiang2025artificial,kirk2024understanding,klein1995characteristics,meincke2024using,park2024generativeagentsimulations1000,peeperkorn2024temperature,radford2019language,reisenbichler2022frontiers,toubia2025twin2k500,vaswani2017attention,wang2023memory,wang2025large,wegner1987paradoxical,wei2022chain,zhang2025verbalized}

\newcommand{\appTemperature}{A}       %
\newcommand{\appTaxonomy}{B}          %
\newcommand{\appModelChoice}{C}       %
\newcommand{\appReplications}{D}      %
\newcommand{\appMeasures}{E}          %
\newcommand{\appCreativePersonas}{F}  %
\newcommand{\appOrdinaryPersonas}{G}  %
\newcommand{\appPromptingStructure}{H}%
\newcommand{\appSeqCot}{I}            %
\newcommand{\appCotPrompts}{J}        %
\newcommand{\appPartitioning}{K}      %

\setlength{\RaggedRightParindent}{0.5in}\RaggedRight

\begin{center}
{\large\bfseries A Cognitive Framework for Idea Diversity in Large Language Models\par}
\vspace{12pt}
Yuting Deng,\quad Melanie Brucks,\quad Olivier Toubia\\[6pt]
{\normalsize Columbia Business School}\\[4pt]
{\small \texttt{\{yd2721, mb4598, ot2107\}@gsb.columbia.edu}}
\end{center}

\primaryheading{Abstract}

Consumers, marketers, and new product developers increasingly rely on LLMs for ideation. Yet ideas generated by independent LLM sessions tend to be less diverse than ideas generated by independent samples of humans, raising concerns that widespread reliance on LLMs could homogenize marketplace ideation and decrease consumer welfare. We identify two mechanisms undermining LLM idea diversity, informed by theories of creative ideation from new product development and cognitive psychology. First, LLMs aggregate knowledge into a unified distribution rather than exhibiting the knowledge partitioning inherent to human populations, where each person occupies a distinct region of the knowledge space. Second, at the individual level, LLMs exhibit fixation just as humans do, where early outputs constrain subsequent ideation. Two product-ideation experiments demonstrate that targeted prompting interventions can address both mechanisms. Persona modifiers based on ordinary individuals improve knowledge partitioning but amplify fixation. This fixation can be mitigated by using chain-of-thought (CoT) prompting, and combining both strategies produces the highest idea diversity, outperforming humans. These findings offer a theoretically grounded framework for understanding LLM idea diversity and practical strategies for leveraging AI's efficiency without compromising the idea diversity that sustains marketplace variety and consumer welfare.


\clearpage

Ask a large language model (LLM) for ten new fitness product ideas and the list will be inventive. LLMs have become powerful ideation tools, capable of generating ideas comparable in novelty and quality to those produced by humans \citep{meincke2024using, boussioux2024crowdless, lee2024empirical, meincke2024diversity, si2024novel}. Accordingly, firms now brainstorm new product offerings with LLMs \citep{dellacqua2023jagged}, marketers use them to generate ad copy \citep{reisenbichler2022frontiers}, and consumer researchers have begun collaborating with them as ideation partners \citep{defreitas2025ideation}. Yet this adoption comes with a hidden cost. Ask again for fitness product ideas in a hundred fresh sessions, and the lists begin to converge, with the same needs served and the same product forms recurring over and over. Researchers have documented this convergence around similar themes and concepts across a range of contexts \citep{defreitas2025ideation, meincke2024diversity} and attribute it to a fundamental limitation of current LLM design, which the computer science literature terms ``mode collapse'' \citep{padmakumar2024does, west2025base,jiang2025artificial}.

LLM idea homogeneity is an urgent problem for consumers, who depend on a marketplace of diverse options. No single offering suits everyone, as consumers' needs, tastes, and constraints vary widely. Further, variety is inherently valuable to consumers, providing agency, novelty, and stimulation \citep{kahn1995consumer, ratner1999choosing}. LLM convergence threatens this variety in two ways. First, the most novel solutions lie far apart in a vast, rugged solution space \citep{meincke2024diversity}, so LLMs that converge on a narrow region leave distant solutions undiscovered. Second, innovation is inherently uncertain \citep{simon1973structure,nickerson2004knowledge}, so consumers benefit when firms place many diverse bets \citep{nelson1961uncertainty,leiponen2010breadth,boudreau2011incentives,boussioux2024crowdless}. When ideation converges, the marketplace's bets concentrate, and fewer successful innovations emerge. Importantly, because idea generation precedes screening, selection, and refinement \citep{hauser2006research, pescher2025role}, it serves as a bottleneck. Downstream development can improve the quality of existing candidates, but it cannot restore diversity that was missing from the start.

This creates a ``tragedy of the commons,'' as researchers and practitioners using LLMs may benefit individually through higher efficiency and originality, while the marketplace as a whole suffers from increased homogeneity \citep{doshi2024evaluating, meincke2024forthcoming, defreitas2025ideation}. We are already seeing this tragedy unfold: a recent analysis of over 41 million research papers shows that using AI tools boosted individual productivity but reduced the diversity of topics explored across the scientific community \citep{hao2026artificial}. 

In this paper, we develop a theoretical account of why LLM ideation converges and how to counteract it. We build on existing research that takes a largely empirical approach to finding prompting strategies that increase output diversity \citep{meincke2024diversity, peeperkorn2024temperature, terwiesch2023large,defreitas2025ideation}. Two strategies show the largest diversity gains \citep{meincke2024diversity}: persona modification, in which the LLM is assigned a specific perspective (e.g., ``you are Steve Jobs,'' or an ``extremely creative persona''), and chain-of-thought prompting, in which the LLM first generates short idea summaries and then explicitly revises them to be bolder and more distinct before finalizing the set. But these strategies were identified empirically, and without a theoretical account of why LLM ideas converge, we lack a principled way to deploy them. For example, it is unclear which personas to select in order to maximize diversity. How to combine strategies is equally uncertain. Interventions that target the same mechanism add cost without adding diversity, whereas interventions that target different mechanisms may be complementary. These uncertainties may have left LLMs' potential in creative tasks unrealized. To our knowledge, no paper has yet leveraged the unique strengths of LLMs to surpass human levels of diversity in idea generation tasks \citep[cf.][]{wang2025large}. 

Prior comparisons of human and LLM diversity also carry two methodological limitations. First, studies rarely match humans and LLMs on what they were asked to do, often giving the model an intervention that human participants never receive. Such comparisons cannot distinguish whether an intervention narrowed the human--LLM gap or would have helped any (human or LLM) ideator given the same instructions. Second, diversity has typically been measured through embedding similarity \citep{meincke2024diversity, doshi2024evaluating}, which is sensitive to wording rather than meaning. Two human ideas with the same meaning can be scored as dissimilar because they are phrased differently, while two LLM ideas with different meanings can be scored as similar because they were phrased similarly by the LLM. Embedding-based measures may therefore overstate the very gap these strategies are meant to close.

We address both limitations here. We test our theoretical account with matched human--LLM comparisons using a meaning-based measure of diversity. In the spirit of recent work suggesting that theories from behavioral science can help explain AI behavior \citep{jackson2025ai}, we draw on work in consumer creativity and cognitive psychology to decompose collective idea diversity into two distinct components: diversity \textit{between} ideators---whether independent sessions explore different regions of the idea space---and diversity \textit{within} ideators---whether a single session explores broadly rather than narrowly. Between-ideator diversity depends on \textit{knowledge partitioning}, which human populations exhibit and LLMs lack, whereas within-ideator diversity is limited by \textit{fixation}, which LLMs exhibit just as humans do. This decomposition explains why existing prompting strategies work and predicts how they interact. We test these predictions in two product-ideation studies and show that combining ordinary personas (targeting partitioning) with chain-of-thought prompting (targeting fixation) produces LLM idea diversity that surpasses human levels.

\primaryheading{The Mechanisms of Convergence}

Foundational research in cognitive psychology points to at least two distinct mechanisms that affect idea diversity among humans, \emph{knowledge partitioning} and \emph{fixation}. Both stem from the fact that every person possesses a unique and limited mental model, a highly personalized associative network built from individual experiences, knowledge, perspectives, and memories \citep{johnsonlaird1983mental, gentner1983structure, moreau2001entrenched}. Mental models play a dual role in creativity. On the one hand, human knowledge is partitioned across billions of individual minds and the idiosyncratic expertise and perspectives of each individual enable diverse exploration at the collective level; on the other hand, a single mental model can become entrenched and constrain individual ideation through fixation. 

In an LLM, analogous mechanisms arise not from a mental model but from three stages of the model's pipeline. During \textit{pre-training}, a neural network is trained on massive text corpora to learn statistical patterns by predicting the next token, with this knowledge encoded in its parameters \citep{vaswani2017attention, brown2020language}. \textit{Post-training} procedures then adapt the pretrained model for instruction following and for alignment with human preferences, typically through supervised fine-tuning (SFT) and Reinforcement Learning from Human Feedback (RLHF) \citep{Christiano2017, Ouyang2022}. Finally, at \textit{inference} time, the model generates text autoregressively, producing each token conditioned on the prompt and preceding context \citep{vaswani2017attention, brown2020language}. 

\begin{figure}[H]
\caption{TWO MECHANISMS OF LLM IDEA CONVERGENCE AND THE INTERVENTIONS THAT TARGET THEM}\label{fig:mechanism-map}

\centering
\includegraphics[width=0.95\textwidth]{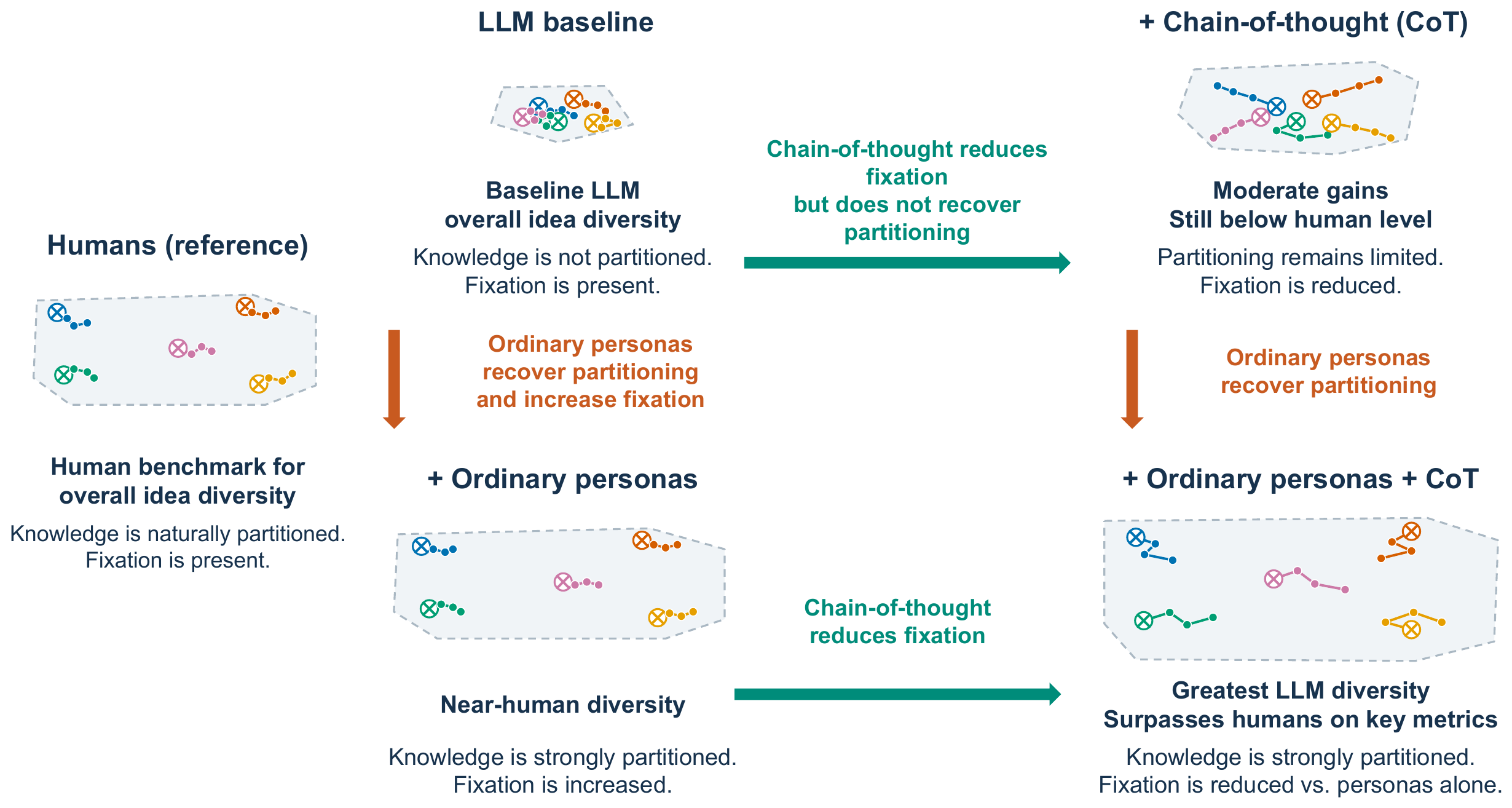}

\end{figure}

As we argue below, these three stages affect both knowledge partitioning and fixation. Figure~\ref{fig:mechanism-map} previews this account, showing the two mechanisms of knowledge partitioning and fixation, where each arises in an LLM, the intervention that targets it, and the predictions we test.

\fignote{An $\times$ marks each session's initial idea, connected dots represent subsequent ideas, and colors distinguish sessions. The dashed region shows the spread of initial ideas. Ordinary personas improve knowledge partitioning but increase fixation (Study~1); chain-of-thought reduces fixation (Study~2). Combining both interventions targets both mechanisms and produces the greatest idea diversity (Study~2).}
\FloatBarrier

\secondaryheading{Knowledge Partitioning}%
Human knowledge and experience are inherently partitioned across people. Each individual's mental model possesses only a tiny, idiosyncratic sample of humankind's total knowledge capital, occupying a unique ``region'' of the possible knowledge space. A birdwatcher, for instance, holds a mental model heavily weighted toward avian species, even though birds represent only an infinitesimal fraction of collective human knowledge.

This partitioning is a key driver of creativity at the group level. When different individuals collectively contribute ideas, their idiosyncratic mental models jointly explore a broader area of the solution space. Crowdsourcing effectively harnesses this variability by combining inputs from heterogeneous individuals \citep{surowiecki2004wisdom, girotra2010idea, leiponen2010breadth, boons2015crowdsourcing}. For example, when tasked with generating new product ideas for fitness, a birdwatcher might propose a hiking app that plans routes around seasonal migration patterns, while a software engineer who enjoys sedentary activities might come up with a fitness app that ``gamifies'' motivational workout plans.

If collective diversity is a function of sampling from many different mental models, LLMs could, in theory, be the ultimate innovators. They are the direct products of massive, digital crowdsourcing of text from millions of different idiosyncratic mental models. But the LLM pipeline does not preserve this partitioning. At the pre-training stage, LLMs minimize cross-entropy loss over a corpus, capturing the aggregate data distribution rather than the distinct distributions that characterize each human \citep{brown2020language, radford2019language}. Post-training alignment mechanisms further erode knowledge partitioning. Specifically, human preference labels are aggregated across annotators, and the model is optimized toward this aggregate signal through processes like RLHF \citep{zhang2025verbalized}. Consequently, LLMs lose knowledge partitioning at both stages. At inference time, they sample from this aggregate probability distribution, producing outputs that reflect central tendencies rather than the idiosyncratic patterns observed in humans.

If aggregation contributes to the reduced diversity often observed in LLM outputs, can we re-partition the model to increase diversity? This is, we argue, what persona modifiers accomplish. Within our theoretical framework, a persona functions as a sampling cue. Providing the model with personas such as ``Zumba-loving college student'' or ``an older woman active in political social media'' can trigger the LLM to navigate toward remote regions of the knowledge space and bypass its centralized distribution. Importantly, this prediction deviates from prior work that suggests using ``creative personas'' such as Elon Musk or Steve Jobs as personas \citep{meincke2024diversity, defreitas2025ideation}. We instead predict that heterogeneous ``ordinary'' personas (ordinary in the sense of being randomly sampled rather than curated around focal traits or notions of creativity) should outperform creative personas. This is because creative personas may be more densely interconnected in the model's knowledge space (frequently co-mentioned and semantically clustered) which limits how far apart the accessed regions can be \citep{wang2025large}. In our framework, whether Elon Musk is actually more creative than any given persona should be irrelevant. For the LLM, a persona is a sampling cue, not a transfer of human cognitive ability. %

\secondaryheading{Fixation}Fixation is a well-documented phenomenon in human creativity, occurring when individuals rely on a dominant mental representation or familiar solution path during idea generation. This tendency can build over time in an idea session, constraining exploration of alternatives \citep{luchins1942mechanization}. Decades of cognitive research have identified two main pathways for how fixation occurs. The first is the part-list cueing effect, where initial activation inhibits the activation of related concepts \citep{slamecka1968partlist, roediger1973output}. %
The second mechanism is unconscious plagiarism, in which external examples activate certain concepts that people unintentionally and unknowingly rely on \citep{johnson1993sourcemonitoring, schacter1999sevensins}. %

Empirical studies across domains provide evidence for fixation in idea generation. For example, product developers primed with benchmarking examples of existing products generate fewer and less novel ideas, as those examples activate specific mental schemas that constrain exploration toward familiar structural forms \citep{smith1991incubation, smith1993constraining, luo2015improving}. Similarly, \citet{bayus2013crowdsourcing} shows that ideators with past success tend to generate less diverse subsequent ideas, because they fixate on the features of their previously implemented ideas. Experts are particularly susceptible to fixation, as their deep knowledge stabilizes their mental models, reducing flexibility compared to novices \citep{dane2010reconsidering, luo2015improving}. This stabilization begins within days of repeated idea generation \citep{brucks2020practice}. %

Because LLMs are trained on massive datasets encompassing a wide range of human perspectives and experiences, they are, by definition, not constrained to a single individual's limited cognitive frame. This might suggest that LLMs should be less prone to fixation. However, we argue that LLMs may exhibit fixation-like behavior despite the diversity of pretraining data. The autoregressive nature of LLM processing is analogous to the associative nature of human mental models. Specifically, because language models generate sequences by recursively predicting the next token based on the preceding context at inference time \citep{vaswani2017attention, brown2020language}, early ideas in a sequence can disproportionately activate related representational patterns. This may lead to repetitive or narrowly focused responses that structurally resemble the initial outputs, analogous to how the associative structure of human mental models causes features from primed examples \citep{dahl2002influence} or previous ideas \citep{dane2010reconsidering, bayus2013crowdsourcing} to unduly influence subsequent activation. For example, when using an LLM to generate a list of fitness product ideas, an early idea like a smart device with progress tracking may anchor subsequent generation on progress tracking. Many of the following ideas may then repeat the same structure, combining sensors with feedback or analytics features, limiting the diversity of the idea set.

Can personas also be used to address fixation? Our framework suggests that if anything, the opposite should occur. Personas enable knowledge partitioning through sampling cues that drop the model off in an idiosyncratic region of the knowledge space. But the persona remains in context throughout generation, and autoregressive conditioning ensures that every subsequent token is generated under the persona's influence. In this sense, the persona acts like a human mental model, and mental models induce fixation. So, then, how can fixation be mitigated? We propose that chain-of-thought (CoT) prompting \citep{wei2022chain} increases within-individual idea diversity because it interrupts fixation. CoT is implemented in idea generation by asking each LLM instance to first generate a set of short idea titles, then explicitly instructing it to modify the titles to ensure they are different, and finally asking it to expand these refined titles into full ideas \citep{meincke2024diversity, defreitas2025ideation}. We hypothesize that this CoT approach counters fixation in two ways. First, requiring short titles rather than full ideas prevents early elaboration, ensuring there are fewer tokens to constrain future output and thereby mitigating token anchoring. Second, explicitly asking the model to make ideas different breaks the typicality bias introduced by RLHF, pushing the model away from the dominant trajectory rather than allowing it to automatically elaborate on the same themes. Put together, CoT should reduce fixation in LLMs by limiting early elaboration and promoting deliberate diversification.

CoT has been studied exclusively as a tool for improving LLM outputs, and to the best of our knowledge, it has never been administered to human ideators. Thus, CoT could be a general aid to idea generation that would help any ideator. However, we suspect that CoT exploits something specific to how LLMs operate. This is because fixation in humans is hard to overturn merely using instructions. For example, when asked to come up with creative ideas, participants in \citet{galinsky2008power} unwittingly borrowed from examples even after being told not to. This difficulty aligns with ``ironic process'' theory. Consciously trying \textit{not} to think about a target representation can paradoxically keep it active and accessible \citep{wegner1987paradoxical}. 

By contrast, LLMs are optimized to follow instructions through post-training procedures such as RLHF \citep{Christiano2017, Ouyang2022}, so they can, in principle, respond to meta-instructions tirelessly and effectively. Although they are not perfect at following instructions \citep{Shi2023, Jang2023}, they may nonetheless implement such check-and-revise loops more effectively than humans. This remains speculative, as prior work has not made this apples-to-apples comparison, and empirically examining whether CoT influences LLMs versus humans differently is a contribution of our work.

\primaryheading{Overview of Studies}

Overall, we test five hypotheses derived from our framework. First, because LLMs lack knowledge partitioning, personas should increase collective diversity by restoring it. Second, ordinary personas, which carry heterogeneous backgrounds and linguistic cues, should outperform creative personas, which carry densely interconnected cues in less distinct regions of the knowledge space. Third, LLMs should exhibit fixation analogous to that of  humans, and personas, by anchoring a session to a single stable perspective, may exacerbate it. Fourth, CoT should mitigate fixation in LLMs (but potentially not in humans). Lastly, because knowledge partitioning and fixation are distinct mechanisms, combining personas with CoT should target both and yield the greatest collective diversity. We test these hypotheses across two studies. 

Unless otherwise specified, idea generation followed the same default procedure for both human and LLM participants across studies. Each participant (human or LLM) independently generated ten ideas for fitness products, submitting each idea sequentially. Human participants received the following prompt:

\begin{tcolorbox}[
  enhanced,
  colback=gray!5!white,
  colframe=gray!50!black,
  title=Prompting Structure for Idea Generation,
  listing only,
  listing engine=listings,
  breakable,
]
\begin{lstlisting}
Enter one idea for a fitness product that would help people improve your fitness here and click `continue' at the bottom of the page. Please enter only one idea here. You will be able to enter additional ideas once you submit this one.
\end{lstlisting}
\end{tcolorbox}

For the LLM simulation, we used GPT-4o (model: \texttt{gpt-4o-2024-11-20}) with default settings (temperature $= 1.0$; Appendix~\appTemperature{} shows that raising the temperature does not close the diversity gap) and structured messages designed to closely mimic the conditions faced by human participants. We replicate the full design and results with additional models (\texttt{gemini-3.1-pro-preview} and the latest reasoning model \texttt{gpt-5.6-sol}; see Appendix~\appModelChoice{} and Appendix~\appReplications{}) to verify that our conclusions do not depend on this choice. Each LLM participant was modeled as a sequence of ten independent API calls, one per idea. To simulate serial ideation, each new call included the model's prior responses in the conversation history, so that the model could ``see'' its previously generated ideas. This mirrored the human condition, where participants generated one idea at a time and could view their previously submitted ideas, and is equivalent to a user in a chat interface asking the model for one idea at a time within a single conversation. For example, the third API call for a simulated LLM participant included the following conversation history:

\begin{tcolorbox}[
  enhanced,
  colback=gray!5!white,
  colframe=gray!50!black,
  title=Prompting Structure for LLM Idea Generation (Default),
  listing only,
  listing engine=listings,
  breakable,
]
\begin{lstlisting}
messages = [
    {"role": "system", "content": "You are a helpful assistant."},
    {"role": "user", "content": "Give me 1 new idea for a new fitness product. The idea should be explained in exactly one sentence. Just give me the idea, labeled `idea #1.'"},
    {"role": "assistant", "content": "**idea #1:** Smart Device that tracks activity..."},
    {"role": "user", "content": "Give me 1 new idea for a new fitness product. The idea should be explained in exactly one sentence. Just give me the idea, labeled `idea #2.'"},
    {"role": "assistant", "content": "**idea #2:** An app that tracks calorie intake..."},
    {"role": "user", "content": "Give me 1 new idea for a new fitness product. The idea should be explained in exactly one sentence. Just give me the idea, labeled `idea #3.'"},
]
\end{lstlisting}
\end{tcolorbox}

\secondaryheading{Measuring Idea Diversity}Previous work comparing human and LLM idea generation often employed embedding-based methods to measure idea similarity. However, these methods are influenced by differences in expression. Human ideas vary in wording even when conveying similar meanings, leading to artificially low similarity scores, while LLM ideas use uniform language even when meanings differ, resulting in artificially high similarity scores. For example, two human ideas that express nearly the same idea, ``eating more fruits and vegetables daily to strengthen the body'' and ``eat plenty of fruit every day, especially apples, oranges, bananas, and cherries,'' have a cosine similarity of only $0.37$. In contrast, two LLM ideas describing different products, a smart \emph{kettlebell} and a smart \emph{jump rope}, have a similarity of $0.71$, largely because both include similar features such as sensors, tracking, and a companion app. Thus, prior studies may have exaggerated the human--LLM gap in idea diversity. To address this limitation, we developed an alternative measure of diversity that does not rely on embeddings. For our measure, we categorize each idea along the three dimensions that define any product idea (industry context, psychological need, and product form) and measure diversity as the breadth of categories an idea set covers. Because categories capture meaning rather than phrasing, two differently worded ideas with the same content are correctly scored as similar, and vice versa (we retain embeddings for two supporting analyses, see Study~1 and Appendix~\appMeasures{}).

\begin{figure}[!tbp]
\caption{CATEGORIZATION STRUCTURE ACROSS THREE DIMENSIONS}\label{fig:categorization}

\centering
\includegraphics[width=\textwidth]{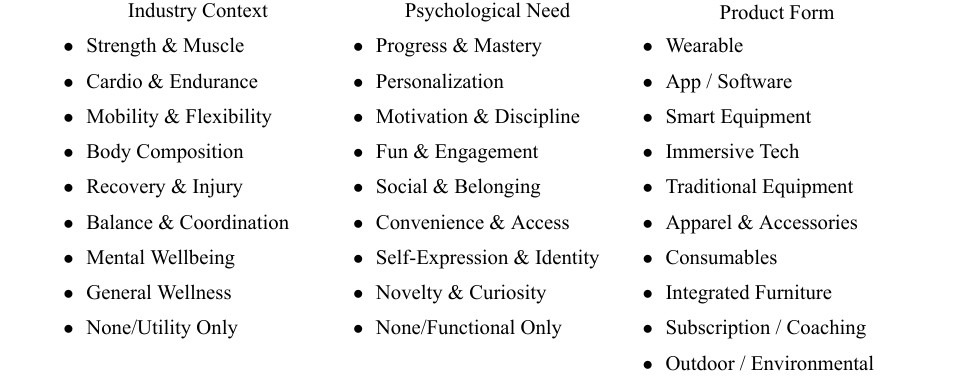}
\end{figure}

\tertiaryheading{Idea Categorization}We first categorized ideas based on content \emph{categories} formed by combining the three dimensions that define any product idea: industry context, psychological need, and product form. We adapted the hierarchical abstraction methodology from \citet{timoshenko2025extract}, who demonstrated that LLMs can effectively organize qualitative data into non-overlapping categories through systematic winnowing and affinitization. Their task involved inferring underlying ``jobs to be done'' from product reviews. Our task of categorizing explicitly stated product ideas is much more straightforward, suggesting that a similar methodology would work well. We therefore first used Claude (model: \texttt{claude-3-5-sonnet-20241022}) to label each idea with detailed subcategories of industry context (e.g., strength, hydration), psychological need (e.g., fun, convenience), and product form (e.g., wearable, app). We then used Claude to consolidate these labels into broader categories following \citeauthor{timoshenko2025extract}'s winnowing procedure, yielding an initial taxonomy of 21 categories (seven per dimension). We reviewed this initial taxonomy, identified categories that overlapped or combined distinct concepts, and refined it into a final taxonomy of 28 non-overlapping categories (9 industry contexts, 9 psychological needs, and 10 product forms); Appendix~\appTaxonomy{} lists the initial themes and the reason for each change. Finally, GPT-4o (model: \texttt{gpt-4o-2024-11-20}; replicated with \texttt{gemini-2.5-flash}, see Appendix~\appModelChoice{}) relabeled every idea using the final taxonomy.
This process yields a categorization structure organized across three dimensions (see Figure~\ref{fig:categorization}). Each fitness product idea is assigned exactly one category from each dimension. As an example, the idea ``a portable dumbbell set that you can take on the road on vacation to keep working out'' would be translated into three categories of Strength \& Muscle (industry context), Convenience \& Access (psychological need), and Traditional Equipment (product form).

\tertiaryheading{Diversity Metrics}We quantified idea diversity using three metrics: \textit{accumulated categories} ($T_{\text{cat}}$), the total number of unique categories covered across all ideas (maximum = 28); \textit{accumulated category combinations} ($T_{\text{comb}}$), the total number of unique combinations across industry context, psychological need, and product form (maximum = $9 \times 9 \times 10 = 810$); and \textit{average pairwise distance} ($d$), the mean number of the three dimensions on which a pair of ideas differs (maximum = 3.0). For example, consider one idea categorized as Cardio \& Endurance (industry context), Progress \& Mastery (psychological need), and Subscription / Coaching (product form), and a second idea categorized as Strength \& Muscle, Progress \& Mastery, and Smart Equipment. For these two ideas, $T_{\text{cat}} = 5$ (five unique categories appear: two industry contexts, one shared psychological need, and two product forms), $T_{\text{comb}} = 2$ (each idea contributes a distinct three-way combination), and $d = 2$ (the ideas differ on industry context and product form but share a psychological need). These metrics capture overall diversity in the full idea set. The next two paragraphs describe how we decompose diversity into its between-ideator and within-ideator components, which correspond to knowledge partitioning and fixation, respectively.

\tertiaryheading{Decomposing Diversity into Partitioning and Fixation} Between-ideator diversity depends on knowledge partitioning, which concerns where ideators \emph{start}. We assess partitioning by computing the three metrics above on each participant's first idea only. The first idea reflects the region of the knowledge space a participant initially retrieves, before any within-session dynamics can operate. Therefore, diversity across first ideas isolates partitioning from fixation. We also illustrate how far apart the semantic neighborhoods accessed by different participants are by embedding every idea and visualizing participant-level centroids.

Within-ideator diversity is limited by fixation, which concerns how exploration unfolds \emph{within} a session. We assess it by tracking the number of unique categories accumulated across a participant's ten sequential ideas and fitting a linear regression to each participant's accumulation curve. The slope $\beta$ represents the participant's exploration rate (maximum = 3.0, as each new idea can add at most one new category per dimension), with flatter slopes indicating greater fixation. 

\tertiaryheading{Robustness}To verify that our findings do not depend on the specific model used, we replicated both the idea generation step with \texttt{gemini-3.1-pro-preview} and \texttt{gpt-5.6-sol}, and the idea categorization step with \texttt{gemini-2.5-flash} in a crossed generator $\times$ categorizer design, holding the prompts, personas, and taxonomy identical. Our findings replicate across models. See Web Appendices~\appModelChoice{} and~\appReplications{}. Additionally, although we favor category-based metrics for the reasons above, the main pattern of results is robust to embedding-based measures. Using the percentage of unique concepts \citep{meincke2024forthcoming} and collective similarity \citep{doshi2024evaluating}, we find the same relative ordering across conditions. However, the embedding-based measures tend to exaggerate the magnitude of the diversity gap between humans and LLMs. For example, the standardized effect size comparing humans with the default LLM under collective diversity is roughly two to three times as large as under the category based measures. Thus, while embedding-based measures capture relative differences across conditions, they may overstate the size of these differences (Appendix~\appMeasures{}).

\primaryheading{Study 1: Knowledge Partitioning and the Human--LLM Diversity Gap}

Study 1 examines the first component of our decomposition, diversity \textit{between} ideators, which our framework attributes to knowledge partitioning. We test whether persona modifiers, acting as sampling cues, can restore the partitioning LLMs lack.

\secondaryheading{Method}%
Study 1 used a four-condition between-participants design. A human condition served as the diversity benchmark. Three LLM conditions varied the persona modifier. The LLM was either given no persona (LLM Default), a persona from a curated set of creative fitness entrepreneurs (LLM Creative Personas), or a persona from a randomly sampled set of ordinary personas (LLM Ordinary Personas).

We recruited $N=121$ human participants from Prolific. Because our design depends on these ideas being generated without the help of LLMs, we excluded participants who appeared to consult external sources during the task. Following \citet{permut2019taskmaster}, we measured the time each participant spent on screens other than the survey and excluded those for whom this exceeded 10\% of the session. After applying this criterion, the final human sample size was $N=99$. 

In the LLM Default condition, we simulated $N=99$ synthetic participants to match the human sample, using the default idea generation procedure described above, with each instance corresponding to an independent participant.

In the LLM Creative Personas and LLM Ordinary Personas conditions, we again use $N = 99$ simulated LLM participants and the same procedure as the LLM Default condition except that each LLM participant was assigned a unique synthetic persona and instructed to generate all ideas from that persona's perspective.
For the Creative Personas, we curated 99 creative personas in the fitness industry using ChatGPT (GPT-5.2). This adapts the creative persona-focused archetype approach from prior work \citep{meincke2024diversity, defreitas2025ideation} to our fitness product context. The set spans a broad range of innovation styles within fitness, from Harpreet Rai, whose Oura Ring made sleep and recovery core fitness metrics, to Chip Wilson, who transformed yoga apparel into a lifestyle brand with Lululemon. %
The complete list is provided in Appendix~\appCreativePersonas{}.
For the Ordinary Personas, we instead draw from the Tencent Personas dataset \citep{ge2024scaling}, which contains over 200{,}000 different synthetic identities representing a broad cross-section of people with various demographics, professions, and life experiences. Here, ``ordinary'' indicates that personas are randomly sampled to capture heterogeneous backgrounds and cues, rather than curated to cluster around fitness-related innovation. We randomly selected 99 distinct personas from the set of 200k publicly shared by \citet{ge2024scaling} to match the number of LLM ideators across conditions (see Appendix~\appOrdinaryPersonas{} for the complete list). Examples include ``a retired politician who has successfully campaigned for library funding,'' ``a newbie software engineer unfamiliar with Sphinx search,'' and ``an award-winning ink chemist who has revolutionized the industry with their inventions.'' Both ordinary personas and creative personas followed the same sequential prompting procedure in Appendix~\appPromptingStructure{}.

\secondaryheading{Results}
We report results in four steps. We first compare overall diversity across the four conditions. We then isolate knowledge partitioning by analyzing first ideas only, and visualize our findings using an embedding analysis of the semantic regions each participant accesses. Finally, we examine fixation by tracking how diversity accumulates within each participant across their ten ideas.

\tertiaryheading{Idea Diversity}  
We first examine the overall diversity of ideas generated by participants in each condition. Consistent with prior research, we find that without any intervention, a group of 99 human participants generate more diverse ideas overall compared to 99 independent LLM sessions (LLM Default). 
Compared to the LLM Default condition, the Human condition produces ideas with more unique categories ($T_{\text{cat}}^{\text{human}} = 28$ vs.\ $T_{\text{cat}}^{\text{llm}} = 23$) and substantially more unique category combinations ($T_{\text{comb}}^{\text{human}} = 208$ vs.\ $T_{\text{comb}}^{\text{llm}} = 89$). We also see that both persona approaches improve idea diversity from the default LLM baseline, with ordinary personas performing best. Creative personas yield 27 categories and 164 unique combinations, a notable improvement though still below human levels. Ordinary personas achieve 27 categories and 208 unique combinations, nearly matching humans in categories (27 vs.\ 28) and matching them in unique combinations (208 vs.\ 208). These results suggest that assigning LLMs personas can elevate LLM diversity to human levels.

\begin{figure}[!htbp]
\caption{STUDY 1 BOOTSTRAPPED DIVERSITY COMPARISONS (10 PARTICIPANTS $\times$ 10 IDEAS)}\label{fig:study3_bootstrap}

\centering
\includegraphics[width=\textwidth]{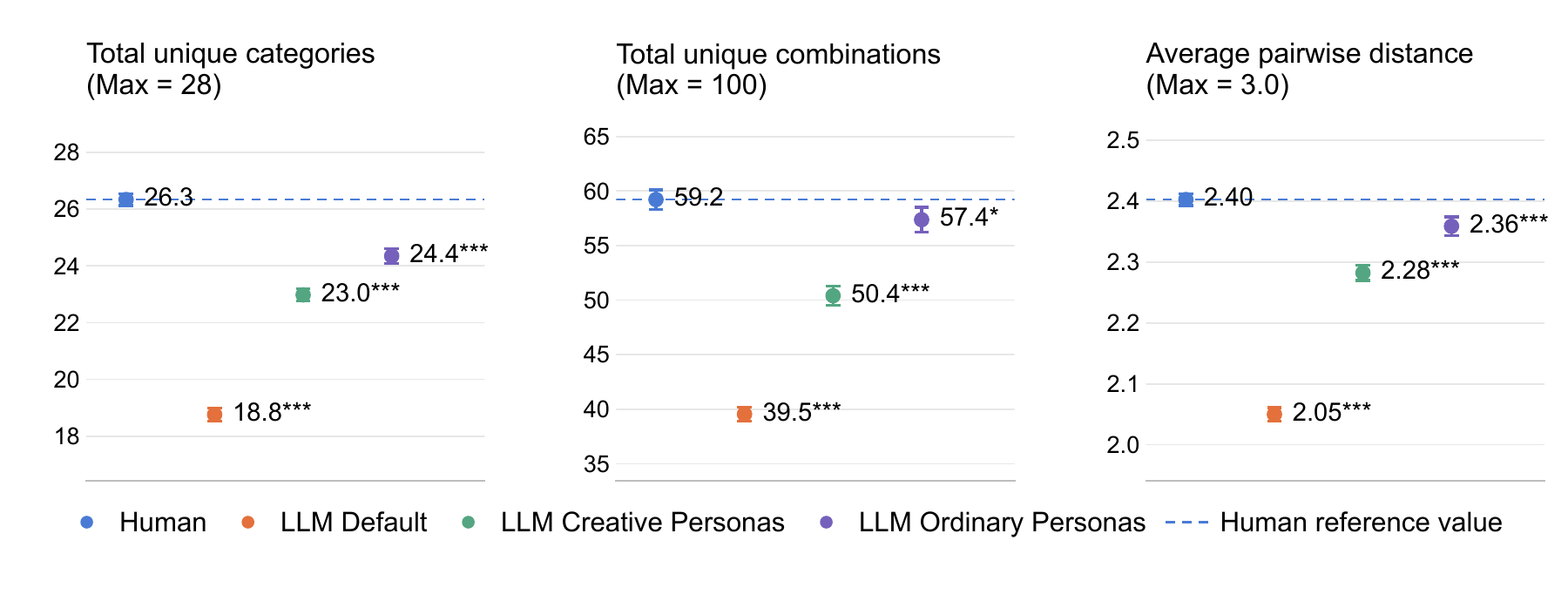}
\fignote{Means from $100$ bootstrap draws of $10$ participants $\times$ $10$ ideas each; error bars are $95\%$ confidence intervals ($\pm 1.96\,SE$). The blue dashed line marks the human reference value. Stars mark Welch $t$-tests on the bootstrap distributions comparing each LLM condition with Human: $^{***}p<.001$, $^{**}p<.01$, $^{*}p<.05$; unmarked comparisons are not significant. Panel titles report theoretical maxima for the sampled idea pool; vertical scales differ across metrics and align the human reference at the same height.}

\end{figure}

Because between-ideator diversity is a property of idea pools across multiple participants, we need nominal groups to statistically evaluate group-level differences. To do this, we bootstrapped samples of 10 humans or 10 LLM agents, each generating 10 ideas (100 ideas total), and repeated this sampling 100 times. Results are shown in Figure~\ref{fig:study3_bootstrap}. Compared to LLM Default, Human groups generate ideas that are significantly more diverse on average, with more unique categories ($T_{\text{cat}}^{\text{human}} = 26.34$, $SE = 0.11$ vs.\ $T_{\text{cat}}^{\text{llm}} = 18.76$, $SE = 0.11$, $p < .001$; maximum = 28 categories), more unique combinations ($T_{\text{comb}}^{\text{human}} = 59.22$, $SE = 0.45$ vs.\ $T_{\text{comb}}^{\text{llm}} = 39.54$, $SE = 0.32$, $p < .001$; 100 ideas can cover at most 100 combinations), and greater pairwise distance ($d^{\text{human}} = 2.40$, $SE = 0.01$ vs.\ $d^{\text{llm}} = 2.05$, $SE = 0.01$, $p < .001$; maximum = 3.0).

Relative to the default LLM, creative personas show significantly higher diversity across all metrics, including unique categories ($T_{\text{cat}}^{\text{creative}} = 22.98$, $SE = 0.11$ vs.\ $T_{\text{cat}}^{\text{default}} = 18.76$, $SE = 0.11$, $p < .001$), unique combinations ($T_{\text{comb}}^{\text{creative}} = 50.40$, $SE = 0.45$ vs.\ $T_{\text{comb}}^{\text{default}} = 39.54$, $SE = 0.32$, $p < .001$), and average pairwise distance ($d^{\text{creative}} = 2.28$, $SE = 0.01$ vs.\ $d^{\text{default}} = 2.05$, $SE = 0.01$, $p < .001$). However, creative personas still fall short of ordinary personas. Ordinary personas generate more unique categories ($T_{\text{cat}}^{\text{ordinary}} = 24.35$, $SE = 0.13$, $p < .001$), more unique combinations ($T_{\text{comb}}^{\text{ordinary}} = 57.37$, $SE = 0.58$, $p < .001$), and higher pairwise distances ($d^{\text{ordinary}} = 2.36$, $SE = 0.01$, $p < .001$) than creative personas. This advantage of ordinary personas over curated creative personas appears to be narrowing in newer models, though it does not reverse. In a full replication with the newer reasoning model \texttt{gpt-5.6-sol}, ordinary personas still lead on the category-based metrics, but by smaller margins (see Appendix~\appReplications{}). This convergence between ordinary and creative personas is the result of both ordinary personas getting worse on first-idea diversity with the newer model, and creative personas becoming better. The latter suggests the newer model inhabits a named founder's specific domain more fully, which also shows in fixation being increased to a level similar to ordinary personas ($\beta = 0.815$ vs.\ $0.816$; see Appendix~\appReplications{}). A persona thus still helps to the extent that it drops a session into a region other sessions do not reach.

We next test whether the two mechanisms in our framework underlie these patterns. 

\begin{figure}[!htbp]
\caption{STUDY 1 KNOWLEDGE PARTITIONING: HUMAN VS.\ LLM CONDITIONS, FIRST IDEAS ONLY}\label{fig:persona_first_ideas}

\centering
\includegraphics[width=\textwidth]{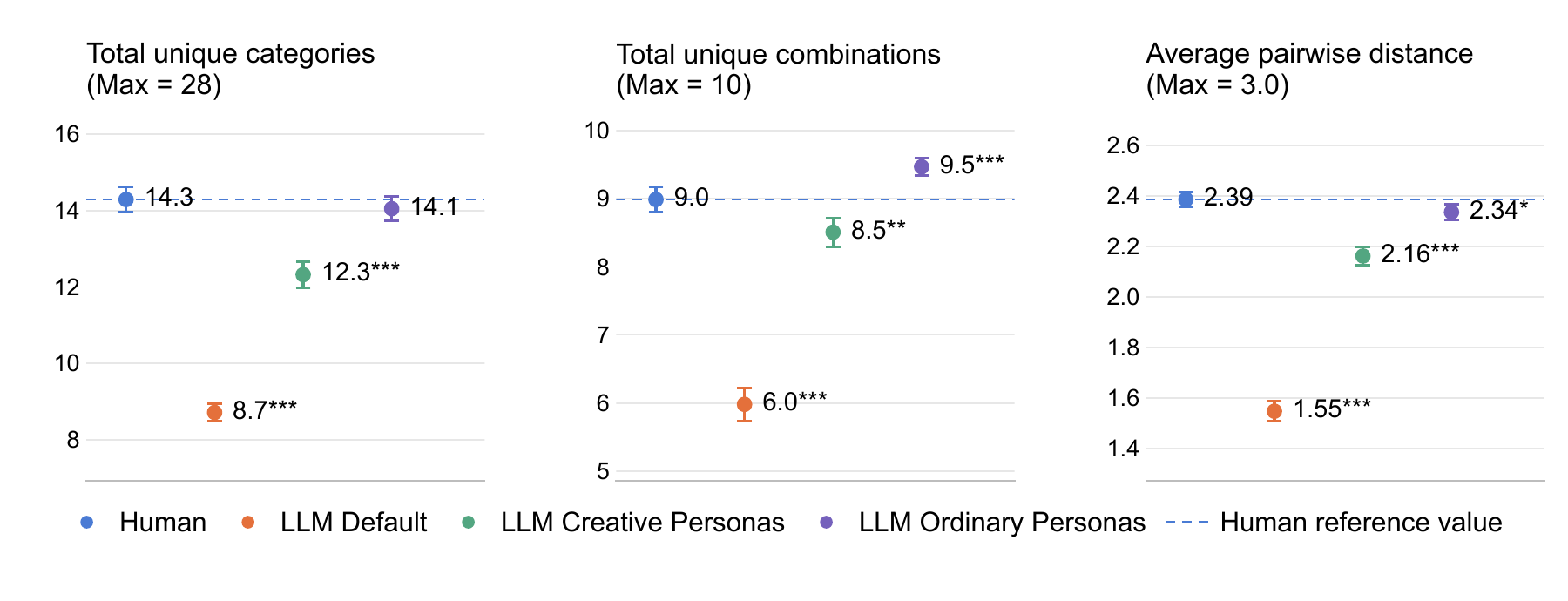}
\fignote{Means from $100$ bootstrap draws of $10$ participants' first ideas each; error bars are $95\%$ confidence intervals ($\pm 1.96\,SE$). The blue dashed line marks the human reference value. Stars mark Welch $t$-tests on the bootstrap distributions comparing each LLM condition with Human: $^{***}p<.001$, $^{**}p<.01$, $^{*}p<.05$; unmarked comparisons are not significant. Panel titles report theoretical maxima for the sampled idea pool; vertical scales differ across metrics and align the human reference at the same height.}

\end{figure}

\tertiaryheading{Knowledge Partitioning (First Ideas)}
 Because knowledge partitioning manifests from the very first moment of retrieval, we analyzed each participant's first idea to measure diversity across participants independent of fixation (which only affects subsequent ideas). Figure~\ref{fig:persona_first_ideas} shows that human participants generate substantially more diverse first ideas than LLM Default participants across all three metrics. Simulated groups of 10 humans produce first ideas with more unique categories ($T_{\text{cat}}^{\text{human}} = 14.29$, $SE = 0.16$ vs.\ $T_{\text{cat}}^{\text{llm}} = 8.71$, $SE = 0.11$, $p < .001$), more unique combinations ($T_{\text{comb}}^{\text{human}} = 8.99$, $SE = 0.10$ vs.\ $T_{\text{comb}}^{\text{llm}} = 5.98$, $SE = 0.12$, $p < .001$), and greater semantic spread in pairwise distances ($d^{\text{human}} = 2.39$, $SE = 0.01$ vs.\ $d^{\text{llm}} = 1.55$, $SE = 0.02$, $p < .001$) versus simulated groups of 10 LLM Default participants. All differences are highly significant, suggesting that humans occupy more distinct regions of the knowledge space from the very beginning, whereas LLMs' first ideas are drawn from a more shared, centralized distribution. Notably, this starting-point collapse is even more severe in newer models. Replicating the design with the more recent reasoning model \texttt{gpt-5.6-sol} roughly doubles the human vs.\ default first-idea gap ($T_{\text{cat}}^{\text{llm}} = 4.22$ there vs.\ $8.71$ here; see Appendix~\appReplications{}), indicating that the partitioning deficit is growing rather than disappearing as models advance. As post-training optimizes a model toward a single preferred response, the activated knowledge region may be narrowing. We return to this in the General Discussion.

LLMs prompted with ordinary personas recover this starting-point gap. Their first ideas cover slightly more unique combinations than humans ($T_{\text{comb}}^{\text{ordinary}} = 9.47$, $SE = 0.07$ vs.\ $T_{\text{comb}}^{\text{human}} = 8.99$, $SE = 0.10$, $p < .001$), match humans on unique categories ($T_{\text{cat}}^{\text{ordinary}} = 14.05$, $SE = 0.16$ vs.\ $T_{\text{cat}}^{\text{human}} = 14.29$, $SE = 0.16$, $p = .30$), and come in slightly below humans on pairwise distance ($d^{\text{ordinary}} = 2.34$, $SE = 0.02$ vs.\ $d^{\text{human}} = 2.39$, $SE = 0.01$, $p = .02$). In comparison, LLM creative personas show significantly lower first-idea diversity than ordinary personas on all three metrics ($T_{\text{cat}}^{\text{creative}} = 12.32$, $SE = 0.17$; $T_{\text{comb}}^{\text{creative}} = 8.51$, $SE = 0.11$; $d^{\text{creative}} = 2.16$, $SE = 0.02$; all $p < .001$ vs.\ ordinary personas). These results suggest that creative persona prompts do not occupy as distinct regions of the idea space as ordinary personas, and thus induce less knowledge partitioning. In the newer \texttt{gpt-5.6-sol}, this partitioning difference shrinks, with the two persona types becoming statistically indistinguishable on first-idea diversity (see Appendix~\appReplications{}).

\FloatBarrier %

\begin{figure}[!htbp]
\caption{STUDY 1 BETWEEN-PARTICIPANT VARIATION, ONE BOOTSTRAP DRAW (t-SNE)}\label{fig:study_3_spread}

\centering
\includegraphics[width=\textwidth]{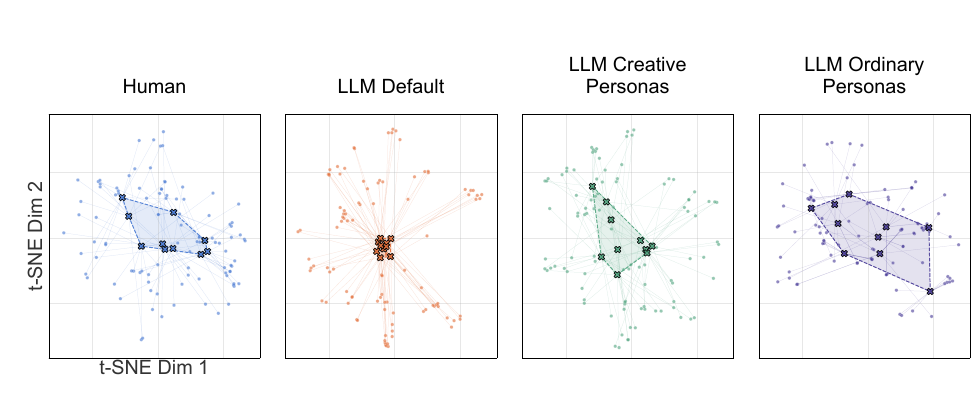}
\fignote{Each panel shows one bootstrap draw of $10$ participants in one condition: dots are ideas, crosses are participant centroids, and the shaded region is the convex hull of the centroids. t-SNE is used for visualization only.}

\end{figure}

\tertiaryheading{Knowledge Partitioning (Embedding Illustration)}
The category-based analyses above show that participants differ in the regions of the idea space from which they retrieve ideas. To illustrate these different semantic neighborhoods, we embedded each idea (OpenAI \texttt{text-embedding-3-small}, 1{,}536 dimensions) and calculated a centroid for each participant (the mean of that participant's idea embeddings). Figure~\ref{fig:study_3_spread} shows, for one bootstrap draw of $10$ participants per condition, the centroids of the $10$ participants, projected into two dimensions with t-SNE \citep{vandermaaten2008visualizing}. The illustration closely mirrors the pattern observed in the category-based analyses. The LLM Default centroids cluster tightly in a central region, whereas Ordinary Personas centroids occupy much more widely separated regions of the semantic space. Human centroids are also broadly dispersed, while Creative Personas centroids span a smaller region than Ordinary Personas.

\begin{figure}[!htbp]
\caption{STUDY 1 FIXATION ANALYSIS: WITHIN-INDIVIDUAL DIVERSITY ACCUMULATION FOR EACH CONDITION}\label{fig:study1_fixation}

\centering
\includegraphics[width=\textwidth]{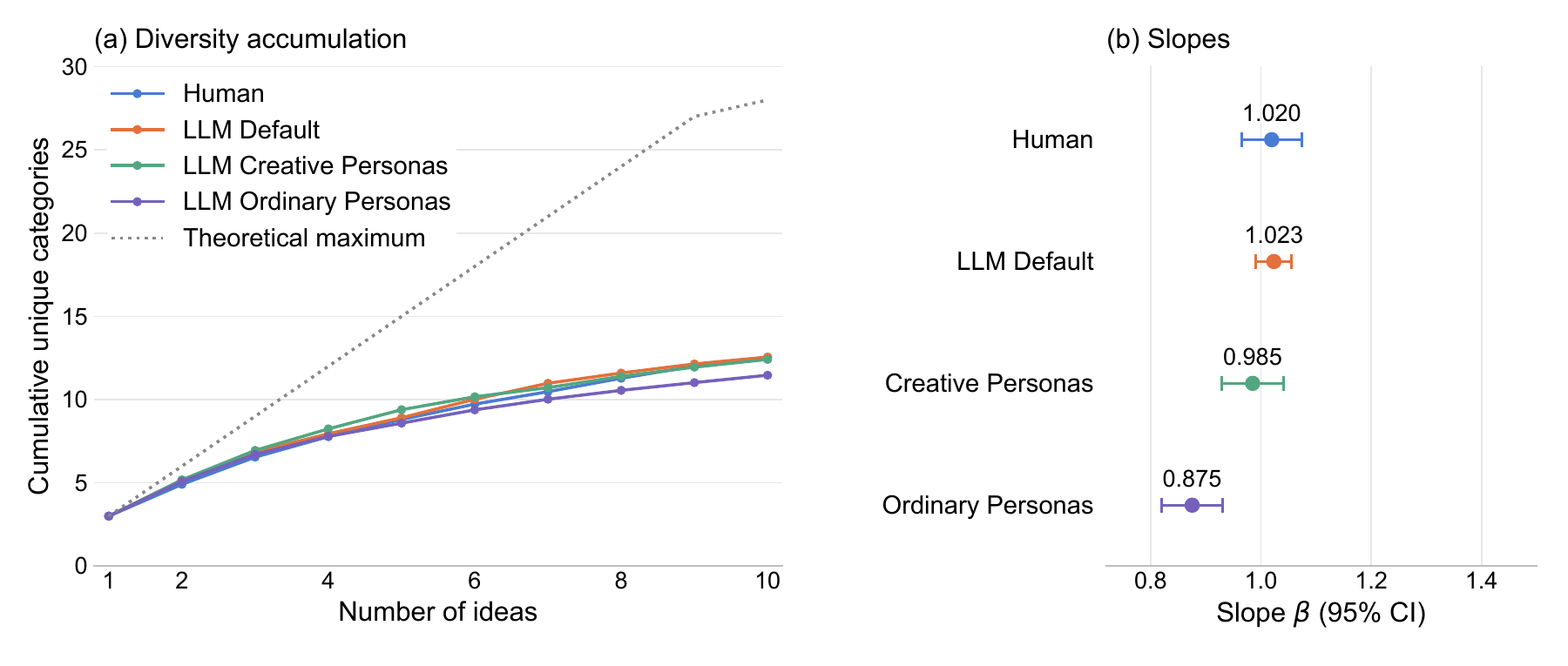}
\fignote{$\beta$ represents the average slope obtained by fitting linear regressions to each participant's diversity accumulation curve. Welch $t$-tests on participant-level slopes: Human vs.\ LLM Default $p=.90$; LLM Default vs.\ Ordinary Personas $p<.001$; LLM Default vs.\ Creative Personas $p=.25$; Ordinary vs.\ Creative Personas $p=.01$. The gray dotted curve shows the theoretical maximum, $\min(3n,28)$ categories after $n$ ideas. The adjacent slope panel shows mean participant-level slopes with $95\%$ confidence intervals ($\pm1.96\,SE$); larger slopes indicate less fixation.}

\end{figure}

\tertiaryheading{Fixation}
To test whether LLMs display ideation behavior akin to fixation, we examined how idea diversity accumulates within each individual across their 10 sequential ideas.
Figure~\ref{fig:study1_fixation} illustrates these accumulation curves. We fit a linear regression to each participant's category accumulation curve to obtain a slope ($\beta$) representing their exploration rate. The maximum potential exploration rate is $\beta = 3.0$, as each new idea can contribute at most three unique categories (one per dimension), with a ceiling of 28 total categories. Human and LLM Default participants accumulate category diversity at nearly identical (and far from the maximal) rates ($\beta_{\text{human}} = 1.020$, $\beta_{\text{llm}} = 1.023$, $p = .90$; Welch two-sample $t$-tests on participant-level slopes, here and throughout the fixation analyses; see Figure~\ref{fig:study1_fixation}). Thus, both LLMs and humans exhibit substantial fixation-like behavior.

As discussed previously, mental models can undermine within-individual diversity through fixation. Thus, by assigning a specific persona to an LLM, are we unwittingly increasing fixation? Our findings suggest that the answer depends on the type of persona. Ordinary personas increase fixation relative to the default condition. The ordinary persona condition exhibits less steep slopes than the default condition for unique categories ($\beta_{\text{default}} = 1.023$, $\beta_{\text{ordinary}} = 0.875$, $p < .001$), indicating greater within-individual fixation. By contrast, creative personas do not change fixation relative to the default condition. Slopes are statistically indistinguishable from the default condition for unique categories ($\beta_{\text{creative}} = 0.985$, $p = .25$). One possible explanation is that ordinary personas provide more concrete, everyday cues that stabilize a participant's exploratory path, whereas creative persona cues are broader and less anchoring, leaving within-participant exploration largely unaffected but limiting between-participant variation. Therefore, while personas can enhance diversity across participants, some persona types (ordinary personas) may simultaneously reinforce a stable schema that constrains exploration within a participant, whereas others (creative personas) do so less.

In sum, Study 1 supports the roles of knowledge partitioning and fixation in the diversity of LLM ideas. It also shows that personas help restore partitioning across participants but can impose fixation within a participant and that this trade-off depends on the persona type.  As partitioning increases through a persona, so does fixation. To address this potential cost, Study 2 turns to CoT prompting, which may counteract fixation by adding an ordered reasoning structure that encourages broader exploration across the solution space. 

\FloatBarrier
\primaryheading{Study 2: Chain-of-Thought Effects on LLM Diversity and Fixation}

Study 1 showed that ordinary personas restore partitioning at the cost of fixation. Study 2 tests whether CoT prompting can remove that cost. CoT prompting encourages broader exploration by breaking the generation process into a series of structured reasoning steps. %
Prior work has applied CoT only to LLMs, leaving open whether the same procedure would help human ideators and thus whether CoT closes a human--LLM gap or simply improves ideation for anyone who follows it. We therefore test CoT in both humans and LLMs. We expect that CoT may benefit LLMs but not humans, reflecting LLMs' unique ability to follow instructions tirelessly and effectively. If so, personas and CoT would target different mechanisms, partitioning and fixation, respectively, and should produce complementary benefits when combined.

\secondaryheading{Method}
Study 2 used a 3 (ideator: Human, LLM Default, LLM Ordinary Personas) $\times$ 2 (procedure: baseline vs. CoT) between-participants design. The three baseline cells are the corresponding conditions from Study 1. The three CoT cells are new: Human CoT, LLM CoT, and LLM Ordinary Personas CoT. 

For the LLM CoT conditions, we followed the chain-of-thought prompt proposed by \citet{defreitas2025ideation}. Specifically, we asked the model to generate all ten ideas \emph{at once} within a single session, and explicitly instructed it to make the revised ideas \emph{bolder} and \emph{more distinct} from one another and to return JSON output only. Of note, this differs from Study 1, where ideas were generated sequentially.\footnote{As a robustness check, we also used the CoT framework with sequential idea generation (same as Study 1) in Appendix~\appSeqCot{}. These results corroborate the main findings from Study 2: CoT improves LLM idea diversity through reducing fixation when applied to LLM Default, and LLM Personas with CoT still yields the highest overall diversity, though the improvement over LLM Personas alone is less pronounced than in the batch setting. } We applied this CoT procedure to two different LLM conditions, each with $N=99$ simulated LLM participants. The only difference was that, in the LLM Ordinary Personas CoT condition, each simulated LLM participant was assigned an ordinary persona (the same set of personas as in Study~1). The original prompt from \citet{defreitas2025ideation} and the LLM CoT and LLM Ordinary Personas CoT prompts we used are in Appendix~\appCotPrompts{}.

For the human CoT condition, we obtained 100 completed survey responses from Prolific participants. As before, to ensure data quality, we excluded five responses whose first idea summary contained 10 characters or fewer and twelve additional responses whose ratio of off-task to on-task time exceeded $0.1$. After applying these criteria, the final sample size was $N = 83$. This condition mirrors the idea generation and revision process used for the LLM CoT conditions. First, participants were asked to generate 10 fitness product ideas all at once on the same page, providing only short summaries rather than full descriptions, as Figure~W2(a) in Appendix~\appCotPrompts{} shows. As with the LLM CoT conditions, this differs from the default setup in Study 1, where participants submitted 10 full ideas sequentially. In the second step, as Figure~W2(b) in Appendix~\appCotPrompts{} shows, participants saw their ten idea summaries from the first step as pre-filled text entries and were explicitly instructed to revise each idea to be as different as possible from their initial version. 

\secondaryheading{Results}
Because total categories and combinations accumulate with the number of participants, comparisons require equal sample sizes. We therefore subsampled the Human and other LLM conditions from Study 1 to the same sample size as the Human CoT condition (the first 83 participants per condition, a deterministic rule fixed before analysis). Aggregating all 83 participants per condition (830 ideas), adding CoT to the default LLM improves diversity, raising total categories from 23 to 27 and unique combinations from 85 to 152. However, this still falls short of human-level performance (196 unique combinations). The LLM Ordinary Personas CoT condition yields the strongest LLM performance, achieving 27 total categories and 240 unique combinations and exceeding humans (vs.\ 196) on the combination metric by a staggering 22\%. At the same time, Human CoT shows a modest \emph{decline} relative to the human baseline, generating the same number of total categories (28 vs.\ 28) but slightly fewer unique combinations (184 vs.\ 196), suggesting that the CoT approach does not necessarily extend to humans and may even slightly constrain their creative exploration. This is consistent with other paradigms in psychology, such as the finding that telling people not to think of a white bear actually increases their likelihood of thinking of a white bear \citep{wegner1987paradoxical}. LLMs, it seems, are immune to this limitation.

\begin{figure}[!htbp]
\caption{STUDY 2 BOOTSTRAPPED DIVERSITY COMPARISON: HUMAN VS.\ LLM (DEFAULT AND CHAIN-OF-THOUGHT)}\label{fig:study4_bootstrap}

\centering
\includegraphics[width=\textwidth]{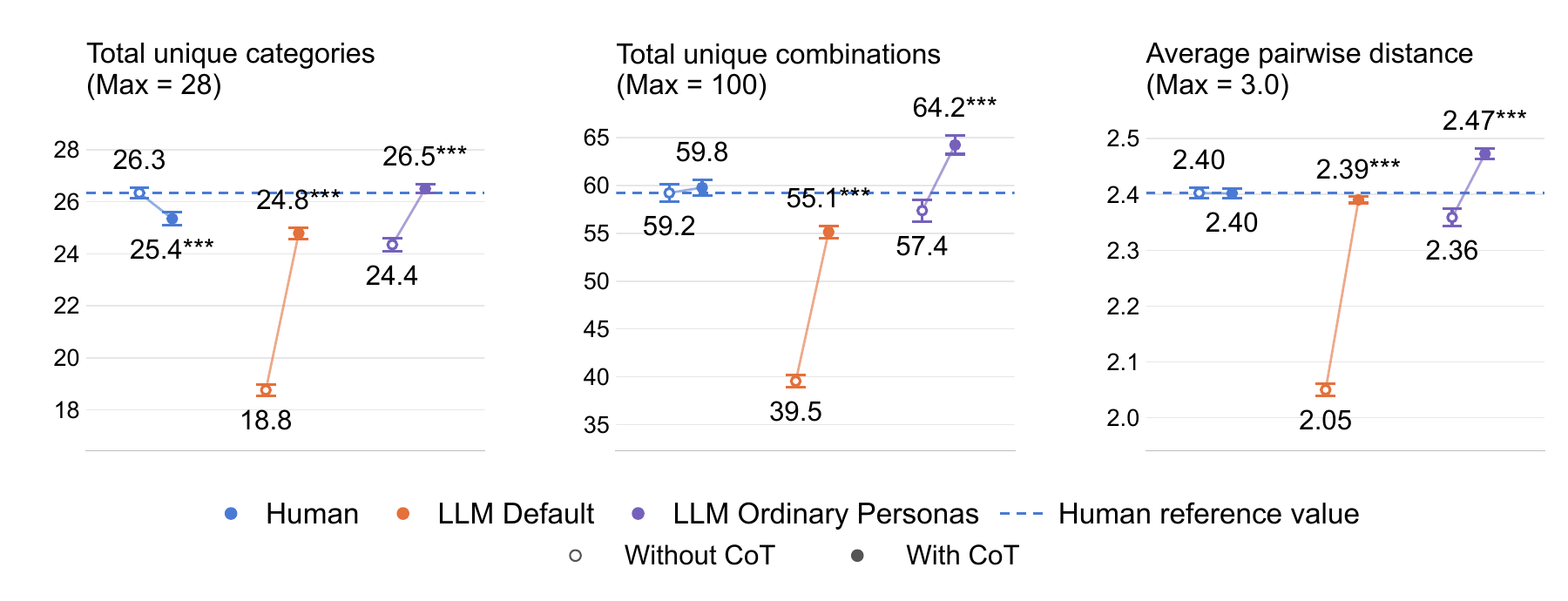}
\fignote{Means from $100$ bootstrap draws of $10$ participants $\times$ $10$ ideas each; error bars are $95\%$ confidence intervals ($\pm 1.96\,SE$). Open markers show baseline conditions and filled markers show the corresponding CoT conditions. Light connecting lines link condition means and do not denote matched participants. The blue dashed line marks the Human baseline reference value. Stars mark Welch $t$-tests on the bootstrap distributions comparing each CoT condition with its baseline (filled vs.\ open marker within each ideator type): $^{***}p<.001$, $^{**}p<.01$, $^{*}p<.05$; unmarked comparisons are not significant. Panel titles report theoretical maxima for the sampled idea pool; vertical scales differ across metrics and align the human reference at the same height.}

\end{figure}

As before, because between-ideator diversity is a property of idea pools across multiple participants, we bootstrap nominal groups to statistically evaluate group-level differences. Bootstrapped analysis further supports these findings and highlights key differences in how CoT affects LLMs and humans (see Figure~\ref{fig:study4_bootstrap} for condition means and confidence intervals). For LLMs, adding CoT to the default model significantly increases all three diversity metrics: unique categories ($SE$s $= 0.11$ and $0.11$, $p < .001$), unique combinations ($SE$s $= 0.32$ and $0.34$, $p < .001$), and pairwise distance ($SE$s $= 0.01$ and $0.00$, $p < .001$). When combined with personas, CoT again increases diversity across all metrics: unique categories ($SE$s $= 0.13$ and $0.09$, $p < .001$), unique combinations ($SE$s $= 0.58$ and $0.49$, $p < .001$), and pairwise distance ($SE$s $= 0.01$ and $0.005$, $p < .001$). For humans, however, CoT does not lead to higher diversity: unique categories decline slightly ($SE$s $= 0.11$ and $0.13$, $p < .001$), while unique combinations ($p = .366$) and pairwise distance ($p = .905$) are statistically unchanged.

\tertiaryheading{Knowledge Partitioning}
Figure~\ref{fig:study_3_spread} illustrates the broader between-participant spread induced by ordinary personas in Study~1. CoT provides no comparable sampling cue to relocate a session's starting point, so our framework predicts that it primarily targets fixation rather than partitioning. The Study~2 first-idea and between-participant analyses are consistent with this prediction (see Appendix~\appPartitioning{} for full results). We therefore turn to fixation.

\begin{figure}[!htbp]
\caption{STUDY 2 FIXATION ANALYSIS: DIVERSITY ACCUMULATION FOR (A) LLMS WITH AND WITHOUT CHAIN-OF-THOUGHT AND PERSONAS; AND (B) HUMANS WITH AND WITHOUT CHAIN-OF-THOUGHT}\label{fig:study4_fixation}

\centering
\includegraphics[width=\textwidth]{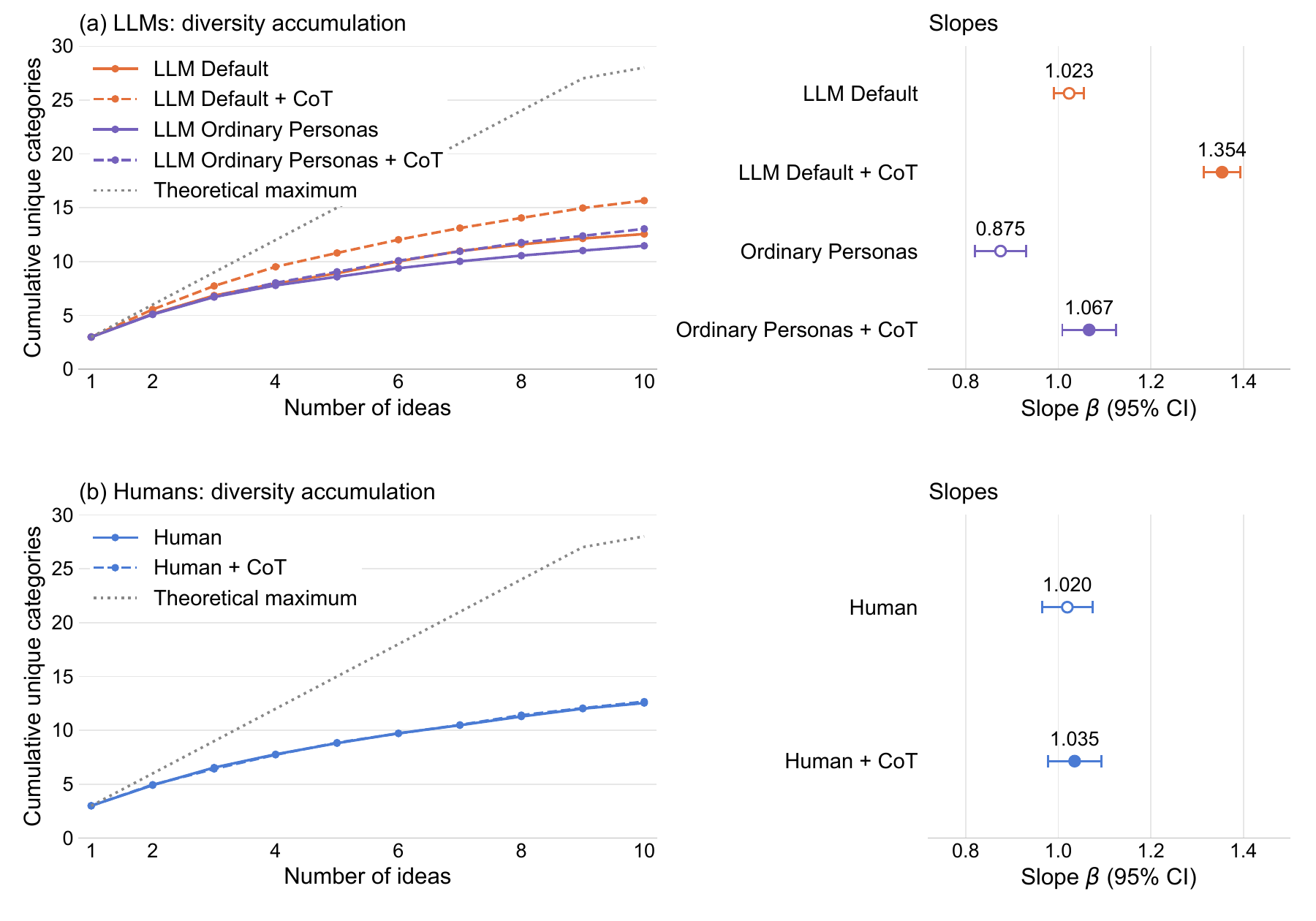}
\fignote{$\beta$ is the average slope from linear regressions on each participant's category accumulation curve; comparisons are Welch $t$-tests on participant-level slopes. Panel~(a): $\beta_{\text{default}} = 1.023$ vs.\ $\beta_{\text{CoT}} = 1.354$ ($p<.001$); $\beta_{\text{ordinary}} = 0.875$ vs.\ $\beta_{\text{ordinary CoT}} = 1.067$ ($p<.001$). Panel~(b): $\beta_{\text{human}} = 1.020$ vs.\ $\beta_{\text{human CoT}} = 1.035$ ($p=.70$). The gray dotted curve shows the theoretical maximum, $\min(3n,28)$ categories after $n$ ideas. The adjacent slope panel shows mean participant-level slopes with $95\%$ confidence intervals ($\pm1.96\,SE$); larger slopes indicate less fixation.}
\end{figure}

\tertiaryheading{Fixation} 
Study 1 shows that personas increase fixation within individual agents by anchoring idea generation to a stable mental model. Here, we test whether CoT can counteract this effect and reduce fixation, both for default LLMs and for persona-based ones. Fixation is again measured as the slope of diversity accumulation across 10 ideas, where steeper slopes indicate broader, less constrained idea exploration and thus lower fixation. As shown in Figure~\ref{fig:study4_fixation}(a), CoT significantly reduces fixation in LLMs. Compared to the Default condition, CoT yields steeper slopes for unique categories ($\beta_{\text{default}} = 1.023$, $SE = 0.017$; $\beta_{\text{CoT}} = 1.354$, $SE = 0.020$; $p < .001$), indicating reduced fixation. Adding CoT to personas also decreases fixation relative to using personas alone, producing steeper slopes (categories: $\beta_{\text{ordinary}} = 0.875$, $SE = 0.028$; $\beta_{\text{ordinary CoT}} = 1.067$, $SE = 0.030$; $p < .001$). In general, these results suggest that CoT is an effective mechanism for LLMs to reduce individual fixation, both alone and paired with personas.
For humans, CoT has no meaningful effect on fixation. As shown in Figure~\ref{fig:study4_fixation}(b), the Human and Human CoT conditions yield nearly identical accumulation slopes for categories ($\beta_{\text{human}} = 1.020$, $SE = 0.028$; $\beta_{\text{human CoT}} = 1.035$, $SE = 0.029$; $p = .697$). This suggests that while CoT offers an external reasoning structure that helps LLMs overcome fixation, humans may not be able to challenge their own fixation in the same way. Once a line of thinking is established, it may be cognitively difficult to deviate from it even with explicit structure.

Importantly, these findings are not specific to any one model. Replicating the design with \texttt{gemini-3.1-pro-preview} and \texttt{gpt-5.6-sol} as generators and \texttt{gemini-2.5-flash} as a second categorizer yields the same ranking of conditions across all generator 
$\times$  categorizer cells (see Appendix~\appModelChoice{}).

\FloatBarrier
\primaryheading{General Discussion}

In this paper, we use theories of consumer creativity and cognitive psychology as analogies for why LLM ideas converge and what can be done about it \citep{jackson2025ai}. Across two studies, we find that LLM diversity is influenced by two mechanisms, knowledge partitioning and fixation. Ordinary personas restored knowledge partitioning across sessions but increased fixation within them. Chain-of-thought prompting reduced fixation but left partitioning unchanged. Combined, the two produced the highest diversity on every metric and exceeded the human benchmark. Figure \ref{fig:mechanism-map} summarizes these results. 

Our findings update best practice for LLM-assisted ideation in consumer research. Consumer researchers now use LLMs to generate hypotheses, stimuli, and synthetic respondents, and current guidelines treat diversity as a matter of prompting the model well \citep{defreitas2025ideation}. Our framework adds a more principled basis for deploying these prompts, along with diagnostics that reveal whether each mechanism is working. In practice this means assigning each session a persona drawn from a heterogeneous set and reporting that set the way one would report a sample, using a chain-of-thought procedure within each session, and examining first-idea diversity for partitioning and within-session accumulation slopes for fixation. 

For firms that brainstorm with LLMs, the framework opens an opportunity beyond off-the-shelf prompting. Ordinary personas already provide a strong baseline, as they are freely available in large public datasets, require no curation, and never underperformed curated creative personas in our data. But firms can go further by writing custom personas that pull sessions toward the distinct use contexts, customer segments, and adjacent categories that matter for their products. A fitness brand, for instance, might assign sessions personas spanning a postpartum mother returning to exercise, a physical therapist, a rural user without gym access, and a competitive powerlifter, each anchoring ideas in a different region of the relevant knowledge space. Our framework gives the design criterion. A persona is valuable to the extent that it drops a session into a region other sessions do not reach, so a portfolio should be judged by the spread of first ideas it produces, not by how elaborate or realistic any single persona is.

Once firms have diversified where sessions begin, the second challenge is keeping each session from narrowing as it unfolds. Here, LLMs offer an advantage specific to machine ideation. Firms can explicitly instruct an LLM to move away from its own emerging patterns in a way that is difficult for human ideators once a line of thought becomes salient. The practical principle is therefore to separate exploration from elaboration by generating distinct directions first, then developing them in depth. However, we also note that the LLM's advantage is far from exhausted when it comes to fixation. Even with personas and CoT combined, sessions accumulate new categories at roughly a third of the theoretical maximum. Thus, there is considerable room for interventions that further loosen the grip of early ideas, and by isolating fixation as a distinct mechanism, our framework gives future work a specific target to aim at.

Our findings speak to consumer welfare as well. Consumers depend on a marketplace of diverse options. Because many firms use the same frontier models with similar default prompts, convergence in model outputs may propagate into convergence in the marketplace itself, reducing the variety of products, services, messages, and experiences available to consumers. Protecting that variety starts with understanding what compromises it. Our framework locates the sources of convergence---lost partitioning across sessions and fixation within them---and in doing so, shows where countermeasures must aim.

That guidance may only grow in importance because the problem shows no sign of resolving itself. The root causes of LLM convergence that we identify (knowledge aggregation during pre-training, RLHF during post-training, and sampling from unified distributions during inference) are inherent to how modern LLMs are trained and deployed, regardless of the specific model. Supporting this, our results replicate across models  (Web Appendices~\appModelChoice{} and~\appReplications{}). Moreover, improvements in general capability may exacerbate the convergence problem, particularly for partitioning. Relative to \texttt{gpt-4o}, \texttt{gpt-5.6-sol} does not improve fixation and its partitioning problem is more severe (Appendix~\appReplications{}). One plausible explanation is that recent post-training's increasing emphasis on reliability, instruction following, and agentic performance may further concentrate models around similar high-probability starting points, while not improving within-session exploration. Regardless of the precise cause, this divergence illustrates the value of distinguishing partitioning from fixation rather than looking at diversity as a whole.  An aggregate diversity score hides which one is getting worse, leaving interventions poorly targeted.

Our framework opens new questions beyond our studies. What sampling cues make the best personas and can cues built from real customer data or digital twins \citep{park2024generativeagentsimulations1000, toubia2025twin2k500} provide richer ones than the synthetic profiles we used? How do partitioning and fixation relate to novelty, feasibility, and usefulness, and how do these mechanisms operate in other creative and problem-solving tasks such as hypothesis generation, strategic scenario planning, and scientific discovery? Whatever the answers, the broader lesson of this framework holds. Convergence arises from distinct mechanisms, and these mechanisms provide a principled approach regarding where to intervene.

\clearpage
\doublespacing

\clearpage
\setcounter{table}{0}\setcounter{figure}{0}
\setlength{\emergencystretch}{3em}
\setcitestyle{aysep={},yysep={;}}
\renewcommand{\primaryheading}[1]{%
  \Needspace{7\baselineskip}%
  \par\addvspace{\baselineskip}%
  {\centering\bfseries\MakeUppercase{#1}\par}%
  \vspace{\baselineskip}%
}

\renewcommand{\secondaryheading}[1]{%
  \par\addvspace{\baselineskip}\noindent #1\par\vspace{\baselineskip}%
}

\renewcommand{\tertiaryheading}[1]{%
  \par\addvspace{\baselineskip}\noindent\hspace*{0.5in}\textit{#1.}\ %
}

\newenvironment{personatable}{%
  \begingroup\singlespacing\fontsize{11.5}{13.8}\selectfont
  \setlength{\tabcolsep}{6pt}\setlength{\extrarowheight}{1pt}
  \setlength{\LTleft}{0pt}\setlength{\LTright}{0pt}
  \setlength{\parindent}{0pt}\setlength{\RaggedRightParindent}{0pt}
}{\endgroup}
\setlist[itemize]{topsep=3pt,itemsep=0pt,parsep=0pt}
\captionsetup[table]{font=samplecaption,labelfont=bf,textfont=normalfont,
  labelsep=period,justification=raggedright,singlelinecheck=false,
  textformat=simple,format=plain,indention=0pt,skip=8pt,position=top}
\renewcommand{\tablename}{Table}

\widowpenalty=10000\clubpenalty=10000
\tcbset{title after break={\textit{Prompt (continued)}}}
\makeatletter\setlength{\@fptop}{0pt}\makeatother

\setlength{\RaggedRightParindent}{0.5in}\RaggedRight\sloppy

\begin{center}
\MakeUppercase{Appendix}
\end{center}

\noindent This appendix provides the temperature analysis (\appTemperature{}), taxonomy development (\appTaxonomy{}), generator--categorizer robustness (\appModelChoice{}), replications with additional models (\appReplications{}), robustness to the diversity measure (\appMeasures{}), creative and ordinary personas (\appCreativePersonas{}--\appOrdinaryPersonas{}), the sequential prompting structure (\appPromptingStructure{}), sequential CoT analyses (\appSeqCot{}), Study 2 chain-of-thought prompts and survey materials (\appCotPrompts{}), and Study 2 knowledge-partitioning analyses (\appPartitioning{}).

\renewcommand{\thetable}{W\arabic{table}}
\renewcommand{\thefigure}{W\arabic{figure}}

\clearpage
\primaryheading{Appendix \appTemperature{}: Temperature Analysis}

We tested temperatures 1.5 and 2.0 with GPT-4o (\texttt{gpt-4o-2024-11-20}), holding other generation settings fixed as in \emph{Idea Generation Procedure}. The reasoning model \texttt{gpt-5.6-sol} (Appendix~\appReplications{}) exposes no temperature parameter. Table~\ref{tab:temperature-sensitivity} uses the temperature-analysis categorization run; values may differ trivially from the current main-text run (Appendix~\appModelChoice{}).

At temperature 2.0, outputs were frequently incoherent or garbled and could not be categorized reliably; diversity metrics are therefore not reported.

Across 990 ideas, increasing temperature from 1.0 to 1.5 raises categories from 22 to 23 and combinations from 89 to 97 (Table~\ref{tab:temperature-sensitivity}), a small increase.

For the same number of ideas, humans reach 28 categories and 209 combinations (Table~\ref{tab:temperature-sensitivity}).

Higher temperature does not close the human--LLM diversity gap and can impair output quality, consistent with \citet{peeperkorn2024temperature}.

\begin{table}[!htbp]
\centering\normalsize\singlespacing\setlength{\tabcolsep}{4pt}\renewcommand{\arraystretch}{1.2}\setlength{\RaggedRightParindent}{0pt}\setlength{\parindent}{0pt}
\caption{\textbf{Temperature Sensitivity Analysis: Diversity Outcomes.} Full samples of $990$ ideas ($99$ participants $\times$ $10$ ideas) per condition; values come from the temperature-analysis categorization run.}\label{tab:temperature-sensitivity}
\begin{tabularx}{\textwidth}{@{}>{\RaggedRight\arraybackslash}X>{\centering\arraybackslash}p{1.20in}>{\centering\arraybackslash}p{1.20in}>{\centering\arraybackslash}p{0.80in}@{}}
\hline\up
\textbf{Group} & \textbf{Total Categories} & \textbf{Unique Combinations} & \textbf{Total Ideas} \\\hline\up
Human & 28 & 209 & 990 \\
LLM Default (Temp = 1.0) & 22 & 89 & 990 \\
LLM Default (Temp = 1.5) & 23 & 97 & 990 \down\\\hline
\end{tabularx}

\end{table}

\clearpage
\primaryheading{Appendix \appTaxonomy{}: Development of the Categorization Taxonomy}

The taxonomy used in \emph{Idea Categorization} was developed in three stages. First, Claude (\texttt{claude-3-5-sonnet-20241022}) labeled each idea with detailed, open-vocabulary subcategories of industry context, psychological need, and product form, producing 49, 106, and 37 distinct labels, respectively. Second, following the winnowing procedure of \citet{timoshenko2025extract}, Claude consolidated these labels into an initial taxonomy of 21 themes (seven per dimension), listed with their definitions in the left columns of Tables~\ref{tab:taxonomy_health}--\ref{tab:taxonomy_form}. Third, we reviewed the initial taxonomy and found that some themes overlapped, some combined distinct concepts, and some frequent kinds of ideas had no theme at all. We revised it into the final taxonomy of 28 categories (9 industry contexts, 9 psychological needs, and 10 product forms). Three decisions applied across dimensions: each idea receives exactly one category per dimension; no residual ``Other'' category is used; and a fourth dimension in the initial scheme, core function, was dropped because it restated the psychological-need dimension. The right columns of Tables~\ref{tab:taxonomy_health}--\ref{tab:taxonomy_form} give the final category for each initial theme and the reason for each change. GPT-4o (\texttt{gpt-4o-2024-11-20}) then assigned every idea one category per dimension using the final definitions; the prompt, the final definitions, and all labeled data are in the deposited materials. All results in the paper use the 28 final categories; the 21 initial themes appear in no analysis.

\begin{table}[!htbp]
\centering\normalsize\singlespacing\setlength{\tabcolsep}{4pt}\renewcommand{\arraystretch}{1.2}\setlength{\RaggedRightParindent}{0pt}\setlength{\parindent}{0pt}
\caption{\textbf{Industry Context: Initial Themes (7) and Final Categories (9).} Definitions in parentheses are those given to the model in the initial taxonomy.}\label{tab:taxonomy_health}
\begin{tabularx}{\textwidth}{@{}>{\RaggedRight\arraybackslash}p{2.05in}>{\RaggedRight\arraybackslash}p{1.35in}>{\RaggedRight\arraybackslash}X@{}}
\hline\up
\textbf{Initial theme (definition)} & \textbf{Final category} & \textbf{Change and reason} \\\hline\up
Strength \& Muscle Development (strength training, muscle building, resistance training) & Strength \& Muscle & Kept; name shortened. \\
Cardiovascular Health (cardio, heart health, endurance) & Cardio \& Endurance & Kept; broadened to aerobic endurance rather than heart health alone. \\
Flexibility \& Mobility (flexibility, mobility, balance, posture) & Mobility \& Flexibility; Balance \& Coordination & Split: range of motion and balance/proprioception are different training goals served by different products (stretching aids vs.\ balance boards). \\
Weight Management (weight loss/gain, body composition, metabolism) & Body Composition & Kept; broadened to fat loss and metabolic goals rather than weight alone. \\
Recovery \& Injury Prevention (rest, recovery, injury prevention, sleep) & Recovery \& Injury & Kept; rehabilitation added. \\
Mental Wellbeing (mental health, stress relief, focus, meditation) & Mental Wellbeing & Kept. \\
Nutrition \& Hydration (diet, nutrition, hydration, supplements) & --- & Dropped as a health goal: it describes what the product is (a consumable) rather than the outcome it serves. These products are coded as \emph{Consumables} on the product-form dimension, with the health goal they target (e.g., Body Composition, Recovery \& Injury, General Wellness). \\
--- & General Wellness & Added: ideas targeting overall fitness with no dominant outcome had no theme. \\
--- & None/Utility Only & Added: peripheral, tool-like products with no clear health impact had no theme. \down\\\hline
\end{tabularx}
\end{table}

\begin{table}[!htbp]
\centering\normalsize\singlespacing\setlength{\tabcolsep}{4pt}\renewcommand{\arraystretch}{1.2}\setlength{\RaggedRightParindent}{0pt}\setlength{\parindent}{0pt}
\caption{\textbf{Psychological Need: Initial Themes (7) and Final Categories (9).}}\label{tab:taxonomy_need}
\begin{tabularx}{\textwidth}{@{}>{\RaggedRight\arraybackslash}p{2.05in}>{\RaggedRight\arraybackslash}p{1.35in}>{\RaggedRight\arraybackslash}X@{}}
\hline\up
\textbf{Initial theme (definition)} & \textbf{Final category} & \textbf{Change and reason} \\\hline\up
Guidance \& Instruction (advice, instruction, form correction, education) & Progress \& Mastery & Merged with Progress Tracking \& Feedback: both describe helping the user improve, and form-correction and tracking products satisfied both definitions, so the boundary was not reliable. \\
Progress Tracking \& Feedback (tracking, feedback, data insights, monitoring) & Progress \& Mastery & Merged (see above). \\
Motivation \& Accountability (motivation, accountability, habit formation) & Motivation \& Discipline & Kept. \\
Convenience \& Time Efficiency (convenience, time management, efficiency) & Convenience \& Access & Merged with Equipment \& Resource Access: portability, space saving, and cost are convenience and access benefits. \\
Equipment \& Resource Access (equipment access, space optimization, portability) & Convenience \& Access & Merged (see above). \\
Social \& Community Connection (social support, community, companionship) & Social \& Belonging & Kept. \\
Personalization \& Customization (personalization, adaptation to individual needs) & Personalization & Kept. \\
--- & Fun \& Engagement & Added: gamification and entertainment, a frequent driver that had no theme. \\
--- & Self-Expression \& Identity & Added: ideas reflecting cultural identity, values, or style had no theme. \\
--- & Novelty \& Curiosity & Added: ideas whose appeal is technological novelty had no theme. \\
--- & None/Functional Only & Added: purely functional products with no psychological component. \down\\\hline
\end{tabularx}
\end{table}

\begin{table}[!htbp]
\centering\normalsize\singlespacing\setlength{\tabcolsep}{4pt}\renewcommand{\arraystretch}{1.2}\setlength{\RaggedRightParindent}{0pt}\setlength{\parindent}{0pt}
\caption{\textbf{Product Form: Initial Themes (7) and Final Categories (10).}}\label{tab:taxonomy_form}
\begin{tabularx}{\textwidth}{@{}>{\RaggedRight\arraybackslash}p{2.05in}>{\RaggedRight\arraybackslash}p{1.35in}>{\RaggedRight\arraybackslash}X@{}}
\hline\up
\textbf{Initial theme (definition)} & \textbf{Final category} & \textbf{Change and reason} \\\hline\up
Wearable Devices (wearables, fitness trackers, smartwatches) & Wearable & Kept. \\
Mobile \& Digital Applications (apps, digital platforms, digital assistants) & App / Software & Kept. \\
Smart Equipment (smart devices, connected fitness equipment) & Smart Equipment & Kept. \\
Traditional Fitness Equipment (non-connected equipment, workout tools) & Traditional Equipment & Kept. \\
Clothing \& Apparel (apparel, footwear, specialized clothing) & Apparel \& Accessories & Broadened to accessories (bags, bottles, straps). \\
Services \& Subscriptions (services, subscriptions, memberships) & Subscription / Coaching & Sharpened to ongoing service delivery (coaching, memberships), so that a coaching app is no longer both an application and a service. \\
Immersive Technologies (VR platforms, VR hardware, AR solutions) & Immersive Tech & Kept. \\
--- & Consumables & Added: receives the supplement, hydration, and nutrition products formerly under Nutrition \& Hydration. \\
--- & Integrated Furniture & Added: furniture-integrated and hybrid-use products had no theme. \\
--- & Outdoor / Environmental & Added: public installations and environment-built products had no theme. \down\\\hline
\end{tabularx}
\end{table}

\clearpage
\primaryheading{Appendix \appModelChoice{}: Robustness to Model Choice (Crossed Generator $\times$ Categorizer Design)}

The main studies use GPT-4o (\texttt{gpt-4o-2024-11-20}) for generation (\emph{Idea Generation Procedure}) and final categorization (\emph{Measuring Idea Diversity}). We varied both roles using the same prompts, personas, and three-dimension taxonomy:
\begin{itemize}
\item Generators: GPT-4o and \texttt{gemini-3.1-pro-preview} (temperature 1.0), and \texttt{gpt-5.6-sol} (a reasoning model with no temperature parameter).
\item Categorizers: GPT-4o or \texttt{gemini-2.5-flash} (temperature 0.3).
\item Samples: $99$ simulated participants $\times$ $10$ ideas per condition; five generator--categorizer cells, with up to $990$ ideas per condition per cell ($956$--$990$ after excluding \textit{none} or out-of-taxonomy labels). Study~1 human ideas provide a reference under both categorizers (Table~\ref{tab:twobytwo_human}).
\end{itemize}

The two categorizers agree substantially at the level of individual labels: Cohen's $\kappa = .704$ overall ($76.9\%$ raw agreement), with per-dimension $\kappa$ ranging from $.641$ (psychological need) to $.748$ (product form).

Across cells (Tables~\ref{tab:twobytwo_tcat}--\ref{tab:twobytwo_d}), the default LLM is least diverse. Personas~$+$~CoT is highest on combinations and distance, and sometimes ties for the most categories. Combination rankings are consistent (mean pairwise Spearman $\rho=.96$; $\rho=1.00$ between Gemini-generated cells), supporting robustness to generator and categorizer choice.

\begin{table}[!htbp]
\centering\normalsize\singlespacing\setlength{\tabcolsep}{4pt}\renewcommand{\arraystretch}{1.2}\setlength{\RaggedRightParindent}{0pt}\setlength{\parindent}{0pt}
\caption{\textbf{Human Reference Under Both Categorizers.}}\label{tab:twobytwo_human}
\begin{tabularx}{\textwidth}{@{}>{\RaggedRight\arraybackslash}X>{\centering\arraybackslash}p{1.20in}>{\centering\arraybackslash}p{1.20in}>{\centering\arraybackslash}p{0.80in}@{}}
\hline\up
\textbf{Categorizer (human ideas)} & $T_{\text{cat}}$ & $T_{\text{comb}}$ & $d$ \\\hline\up
\texttt{gpt-4o-2024-11-20}                    & 28 & 208 & 2.426 \\
\texttt{gemini-2.5-flash} & 28 & 224 & 2.439 \down\\\hline
\end{tabularx}

\end{table}

\begin{table}[!htbp]
\centering\normalsize\singlespacing\setlength{\tabcolsep}{4pt}\renewcommand{\arraystretch}{1.2}\setlength{\RaggedRightParindent}{0pt}\setlength{\parindent}{0pt}
\caption{\textbf{Accumulated Categories $T_{\text{cat}}$ (max $= 28$) by Generator--Categorizer Cell.} Notes apply to Tables~\ref{tab:twobytwo_human}--\ref{tab:twobytwo_d}. The GPT-gen\,/\,GPT-cat cell is the main-text basis (the \emph{Idea Generation Procedure} section); it re-applies the categorization pipeline to the main-study ideas, so its values can differ trivially across categorization runs. The Sol-gen cell uses ideas generated by the newer reasoning model \texttt{gpt-5.6-sol}, examined in detail in Appendix~\appReplications{}.}\label{tab:twobytwo_tcat}
\begin{tabular}{@{}L{1.65in}C{0.85in}C{0.85in}C{0.85in}C{0.85in}C{0.85in}@{}}
\hline\up
\textbf{Condition} & \textbf{GPT-gen / GPT-cat} & \textbf{Sol-gen / GPT-cat} & \textbf{GPT-gen / Gem-cat} & \textbf{Gem-gen / GPT-cat} & \textbf{Gem-gen / Gem-cat} \\\hline\up
LLM Default                & 23 & 19 & 24 & 18 & 18 \\
LLM CoT                    & 27 & 25 & 28 & 23 & 24 \\
LLM Creative Personas & 27 & 27 & 27 & 27 & 27 \\
LLM Ordinary Personas       & 27 & 27 & 28 & 28 & 28 \\
LLM Personas $+$ CoT        & 28 & 27 & 28 & 28 & 28 \down\\\hline
\end{tabular}

\end{table}

\begin{table}[!htbp]
\centering\normalsize\singlespacing\setlength{\tabcolsep}{4pt}\renewcommand{\arraystretch}{1.2}\setlength{\RaggedRightParindent}{0pt}\setlength{\parindent}{0pt}
\caption{\textbf{Accumulated Category Combinations $T_{\text{comb}}$ (max $= 810$) by Generator--Categorizer Cell.}}\label{tab:twobytwo_tcomb}
\begin{tabular}{@{}L{1.65in}C{0.85in}C{0.85in}C{0.85in}C{0.85in}C{0.85in}@{}}
\hline\up
\textbf{Condition} & \textbf{GPT-gen / GPT-cat} & \textbf{Sol-gen / GPT-cat} & \textbf{GPT-gen / Gem-cat} & \textbf{Gem-gen / GPT-cat} & \textbf{Gem-gen / Gem-cat} \\\hline\up
LLM Default                & 89  & 68  & 90  & 63  & 58  \\
LLM CoT                    & 162 & 144 & 167 & 135 & 109 \\
LLM Creative Personas & 164 & 172 & 165 & 196 & 202 \\
LLM Ordinary Personas       & 208 & 198 & 227 & 215 & 219 \\
LLM Personas $+$ CoT        & 264 & 263 & 283 & 274 & 281 \down\\\hline
\end{tabular}

\end{table}

\begin{table}[!htbp]
\centering\normalsize\singlespacing\setlength{\tabcolsep}{4pt}\renewcommand{\arraystretch}{1.2}\setlength{\RaggedRightParindent}{0pt}\setlength{\parindent}{0pt}
\caption{\textbf{Average Pairwise Distance $d$ (max $= 3.0$) by Generator--Categorizer Cell.}}\label{tab:twobytwo_d}
\begin{tabular}{@{}L{1.65in}C{0.85in}C{0.85in}C{0.85in}C{0.85in}C{0.85in}@{}}
\hline\up
\textbf{Condition} & \textbf{GPT-gen / GPT-cat} & \textbf{Sol-gen / GPT-cat} & \textbf{GPT-gen / Gem-cat} & \textbf{Gem-gen / GPT-cat} & \textbf{Gem-gen / Gem-cat} \\\hline\up
LLM Default                & 2.046 & 1.902 & 1.871 & 1.778 & 1.864 \\
LLM CoT                    & 2.380 & 2.407 & 2.319 & 2.253 & 2.081 \\
LLM Creative Personas & 2.304 & 2.334 & 2.238 & 2.363 & 2.345 \\
LLM Ordinary Personas       & 2.401 & 2.302 & 2.375 & 2.321 & 2.312 \\
LLM Personas $+$ CoT        & 2.509 & 2.515 & 2.505 & 2.461 & 2.434 \down\\\hline
\end{tabular}

\end{table}

\clearpage
\primaryheading{Appendix \appReplications{}: Replications with Additional Models (\texttt{gemini-3.1-pro-preview} and \texttt{gpt-5.6-sol})}

We replicated Studies~1 and~2 with OpenAI's \texttt{gpt-5.6-sol} (accessed August 2026; default sampling, low reasoning effort, no temperature parameter), using the original prompts, personas, and GPT-4o categorization pipeline. This is the Sol-gen\,/\,GPT-cat cell in Appendix~\appModelChoice{}. Table~\ref{tab:sol_summary} follows the main-text reporting format.

\begin{table}[!htbp]
\centering\normalsize\singlespacing\setlength{\tabcolsep}{4pt}\renewcommand{\arraystretch}{1.2}\setlength{\RaggedRightParindent}{0pt}\setlength{\parindent}{0pt}
\caption{\textbf{Key Statistics for the \texttt{gpt-5.6-sol} Replication.} Same procedures as the main text: full-sample values over up to $990$ ideas per condition ($966$--$990$ after excluding ideas assigned \textit{none} or out-of-taxonomy labels); bootstrapped means over $100$ draws of $10$ participants $\times$ $10$ ideas; first-idea $T_{\text{cat}}$ over $100$ draws of $10$ participants' first ideas; $\beta$ is the mean participant-level category accumulation slope. The Human row is the same human data as the main text.}\label{tab:sol_summary}
\begin{tabular}{@{}L{1.55in}C{0.59in}C{0.59in}C{0.59in}C{0.59in}C{0.59in}C{0.59in}C{0.59in}@{}}
\hline\up
& \multicolumn{2}{c}{\textbf{Full sample}} & \multicolumn{3}{c}{\textbf{Bootstrapped ($10 \times 10$)}} & \textbf{First ideas} & \textbf{Fixation} \\
\textbf{Condition} & $T_{\text{cat}}$ & $T_{\text{comb}}$ & $T_{\text{cat}}$ & $T_{\text{comb}}$ & $d$ & $T_{\text{cat}}$ & $\beta$ \\\hline\up
Human (reference)          & 28 & 208 & 26.34 & 59.22 & 2.40 & 14.29 & 1.020 \\
LLM Default                & 19 & 68  & 16.28 & 31.00 & 1.91 & 4.22  & 0.992 \\
LLM CoT                    & 25 & 144 & 23.21 & 52.37 & 2.41 & 8.25  & 1.410 \\
LLM Creative Personas & 27 & 172 & 21.89 & 50.68 & 2.29 & 12.99 & 0.815 \\
LLM Ordinary Personas       & 27 & 198 & 22.71 & 53.27 & 2.26 & 13.05 & 0.816 \\
LLM Personas $+$ CoT        & 27 & 263 & 26.05 & 68.12 & 2.49 & 15.10 & 1.215 \down\\\hline
\end{tabular}

\end{table}

Four findings stand out. First, every effect in the paper replicates in direction: the default LLM is the least diverse condition on every metric, both persona types and CoT improve diversity, CoT reduces fixation ($\beta = 0.992 \to 1.410$) while personas increase it, and LLM Personas~$+$~CoT is the most diverse condition overall, again exceeding the human benchmark on unique combinations.

\tertiaryheading{Baseline Diversity} Default first-idea diversity is $T_{\text{cat}}=4.22$ ($SE=0.09$), versus $8.71$ for GPT-4o and $9.14$ for \texttt{gemini-3.1-pro-preview} under identical prompts, roughly doubling the human--default starting gap. Full-sample combinations fall from $89$ (GPT-4o) to $68$, and between-participant variation ($M=0.225$) is lowest among the three generators. 

\tertiaryheading{Ordinary versus Creative Personas} Ordinary personas outperform creative personas on all three GPT-4o bootstrap metrics ($p<.001$). On \texttt{gpt-5.6-sol}:
\begin{itemize}
\item Categories: $22.71$ vs.\ $21.89$, $p<.001$ (GPT-4o: $24.35$ vs.\ $22.98$). Combinations: $53.27$ vs.\ $50.68$, $p<.001$ (GPT-4o: $57.37$ vs.\ $50.40$).
\item The full-sample combination gap narrows from $44$ to $26$ (session-pool permutation test, $p=.07$); first-idea diversity does not differ significantly.
\item Creative personas lead on distance: $d=2.29$ vs.\ $2.26$, $p=.019$.
\item Both persona types increase fixation similarly ($\beta=0.815$ vs.\ $0.816$, both $p<.001$ vs.\ default); creative personas left fixation unchanged on GPT-4o.
\end{itemize}

\tertiaryheading{Why Persona Types Converge in Newer Models} Our framework suggests one possible interpretation of this pattern. In GPT-4o, ordinary personas outperform creative personas both in overall diversity and in first-idea diversity, consistent with heterogeneous ordinary personas providing more distinct sampling cues. In \texttt{gpt-5.6-sol}, ordinary personas still produce greater overall diversity, but their advantage is smaller, and the two persona types become statistically indistinguishable on first-idea diversity. This suggests that the distinction between persona types may matter less for where newer models begin, even though it can still matter for the diversity of the full idea set. One possibility is that stronger instruction following makes newer models more responsive to persona prompts in general, reducing the difference between heterogeneous ordinary cues and curated creative ones. This interpretation remains speculative, but the broader result is clear: personas continue to improve diversity, while the incremental benefit of choosing ordinary over creative personas appears to shrink as models advance.

\tertiaryheading{Human Comparison} Ordinary personas recover most first-idea diversity but remain below humans ($T_{\text{cat}}=13.05$ vs.\ $14.29$, $p<.001$); on GPT-4o, this difference is not significant.

\tertiaryheading{Replication with \texttt{gemini-3.1-pro-preview}}
We replicated Studies~1 and~2 with \texttt{gemini-3.1-pro-preview}, using the original prompts, personas, and GPT-4o categorizer (Gemini-gen\,/\,GPT-cat in Appendix~\appModelChoice{}). Table~\ref{tab:gemini_summary} follows Table~\ref{tab:sol_summary}.

\tertiaryheading{Core Results} Default is least diverse on every full-sample and bootstrap metric. Ordinary personas reach $28$ categories and $215$ combinations (human: $208$); they outperform creative personas on bootstrap categories ($23.37$ vs.\ $22.72$, $p=.008$) and combinations ($51.03$ vs.\ $48.86$, $p=.005$), and match human first-idea diversity ($14.11$ vs.\ $14.29$, $p=.44$). CoT reduces fixation ($\beta=0.895\to1.204$, $p<.001$), whereas personas increase it ($\beta=0.680$ and $0.641$). Personas~$+$~CoT is most diverse overall ($274$ full-sample combinations).

\tertiaryheading{Differences from GPT-4o} CoT narrows first-idea diversity ($T_{\text{cat}}=7.11$ vs.\ default $9.14$, $p<.001$). Default is more fixated than humans ($\beta=0.895$ vs.\ $1.020$, $p<.001$), unlike the nonsignificant GPT-4o difference. Creative personas exceed ordinary personas on distance ($d=2.31$ vs.\ $2.25$, $p<.001$), also observed for \texttt{gpt-5.6-sol}.

\begin{table}[!htbp]
\centering\normalsize\singlespacing\setlength{\tabcolsep}{4pt}\renewcommand{\arraystretch}{1.2}\setlength{\RaggedRightParindent}{0pt}\setlength{\parindent}{0pt}
\caption{\textbf{Key Statistics for the \texttt{gemini-3.1-pro-preview} Replication.} Same procedures and format as Table~\ref{tab:sol_summary}: full-sample values over up to $990$ ideas per condition ($956$--$990$ after excluding ideas assigned \textit{none} or out-of-taxonomy labels); bootstrapped means over $100$ draws of $10$ participants $\times$ $10$ ideas; first-idea $T_{\text{cat}}$ over $100$ draws of $10$ participants' first ideas; $\beta$ is the mean participant-level category accumulation slope. The Human row is the same human data as the main text.}\label{tab:gemini_summary}
\begin{tabular}{@{}L{1.55in}C{0.59in}C{0.59in}C{0.59in}C{0.59in}C{0.59in}C{0.59in}C{0.59in}@{}}
\hline\up
& \multicolumn{2}{c}{\textbf{Full sample}} & \multicolumn{3}{c}{\textbf{Bootstrapped ($10 \times 10$)}} & \textbf{First ideas} & \textbf{Fixation} \\
\textbf{Condition} & $T_{\text{cat}}$ & $T_{\text{comb}}$ & $T_{\text{cat}}$ & $T_{\text{comb}}$ & $d$ & $T_{\text{cat}}$ & $\beta$ \\\hline\up
Human (reference)          & 28 & 208 & 26.34 & 59.22 & 2.40 & 14.29 & 1.020 \\
LLM Default                & 18 & 63  & 16.38 & 29.43 & 1.79 & 9.14  & 0.895 \\
LLM CoT                    & 23 & 135 & 21.39 & 52.04 & 2.26 & 7.11  & 1.204 \\
LLM Creative Personas & 27 & 196 & 22.72 & 48.86 & 2.31 & 14.11 & 0.641 \\
LLM Ordinary Personas       & 28 & 215 & 23.37 & 51.03 & 2.25 & 14.11 & 0.680 \\
LLM Personas $+$ CoT        & 28 & 274 & 26.48 & 62.71 & 2.42 & 15.03 & 0.995 \down\\\hline
\end{tabular}

\end{table}

\clearpage
\primaryheading{Appendix \appMeasures{}: Robustness to the Diversity Measure}

We applied two published embedding measures to GPT-4o (main-text basis), \texttt{gemini-3.1-pro-preview}, and \texttt{gpt-5.6-sol} ideas:
\begin{itemize}
\item \emph{Method 1} \citep{meincke2024forthcoming}: percentage of unique concepts. Ideas are linked when embedding cosine similarity exceeds $\theta$; connected components define concepts. We use the original Universal Sentence Encoder \citep{cer2018universal} and $\theta=.80$.
\item \emph{Method 2} \citep{doshi2024evaluating}: each idea's cosine similarity to the mean embedding of all other ideas in its condition, scaled $0$--$100$, using the original \texttt{text-embedding-ada-002}. We report $100$ minus similarity, so higher values mean greater diversity.
\end{itemize}

Table~\ref{tab:embedding_robustness} reports both measures. Both embedding-based methods reproduce the pattern from the main text: the default LLM yields the least diverse idea pool, LLM Personas~$+$~CoT the most diverse among LLM conditions, and the human pool sits near the top under Method~1 ($98.8\%$ unique concepts, above every LLM condition except LLM Personas~$+$~CoT) and at the top under Method~2 (collective diversity $11.6$, the highest value observed). For \texttt{gpt-5.6-sol}, default and Personas~$+$~CoT are respectively lowest and highest among LLMs (Method~1: $36.6\%$ and $100\%$; Method~2: $7.6$ and $10.1$). Across generators, Method~1 rankings agree at pairwise Spearman $\rho\geq.97$; Method~2 rankings agree at $\rho=.70$--$.90$, with CoT's middle rank changing (Gemini CoT is more homogeneous than GPT CoT). The least and most diverse conditions are identical across generators and methods.

Gap magnitudes depend on both the measure and the pool size. In the full $990$-idea pools, Method~1 yields $47.8\%$ unique concepts for the default GPT-4o condition versus $98.8\%$ for humans, while Method~2 yields collective diversity of $6.9$ versus $11.6$. These raw gaps have different units. To compare standardized separation on a common sampling unit, we applied all five measures to the same $100$ draws of $10$ participants with $10$ ideas each from the human and default GPT-4o conditions (Table~\ref{tab:standardized_measure_gaps}). The descriptive standardized human--default gaps are $4.59$ for Method~1 and $16.40$ for Method~2, compared with $5.03$--$6.82$ for the three category-based measures. Method~2 therefore yields a standardized gap about $2.4$--$3.3$ times as large as the category-based gaps, supporting the concern that collective similarity magnifies the apparent human--LLM difference. Method~1 does not: its standardized gap is smaller than all three category-based gaps. Because these metrics capture different aspects of diversity, the comparison does not establish a single true gap or an inflation factor. We use category metrics to characterize substantive diversity and embeddings as a robustness check on the broad condition pattern; embedding similarity can conflate wording and meaning, and its values depend on the encoder and threshold.

\begin{table}[!htbp]
\centering\normalsize\singlespacing\setlength{\tabcolsep}{4pt}\renewcommand{\arraystretch}{1.2}\setlength{\RaggedRightParindent}{0pt}\setlength{\parindent}{0pt}
\caption{\textbf{Embedding-Based Collective Diversity by Method and Generator.} $N = 990$ ideas per condition per generator. The human pool is generator-independent, so a single reference value is shown per method.}\label{tab:embedding_robustness}
\begin{tabular}{@{}L{1.55in}C{0.70in}C{0.70in}C{0.70in}C{0.70in}C{0.70in}C{0.70in}@{}}
\hline\up
& \multicolumn{3}{c}{\textbf{Method 1: \% unique ($\theta=.80$)}} & \multicolumn{3}{c}{\textbf{Method 2: diversity (100 $-$ sim)}} \\
\textbf{Condition} & \textbf{GPT-gen} & \makecell{\textbf{Gemini-}\\\textbf{gen}} & \textbf{Sol-gen} & \textbf{GPT-gen} & \makecell{\textbf{Gemini-}\\\textbf{gen}} & \textbf{Sol-gen} \\\hline\up
LLM Default                & 47.8\% & 56.1\% & 36.6\% & 6.9  & 8.1  & 7.6  \\
LLM CoT                    & 78.0\% & 94.0\% & 75.1\% & 9.3  & 8.8  & 9.5  \\
LLM Creative Personas & 84.5\% & 99.9\% & 99.2\% & 7.6  & 9.3  & 9.2  \\
LLM Ordinary Personas       & 97.9\% & 100.0\% & 99.6\% & 8.5  & 10.6 & 9.9  \\
LLM Personas $+$ CoT        & 100.0\% & 100.0\% & 100.0\% & 10.1 & 11.3 & 10.1 \\
Human (reference)          & \multicolumn{3}{c}{98.8\%} & \multicolumn{3}{c}{11.6} \down\\\hline
\end{tabular}

\end{table}

\begin{table}[!tbp]
\centering
\caption{\textbf{Human--Default LLM Effects in Matched 100-Idea Pools.} Pool-level means and standard deviations (SD) from $100$ draws per condition.}\label{tab:standardized_measure_gaps}
\resizebox{\textwidth}{!}{%
\begin{tabular}{@{}lcccc@{}}
\hline\up
\textbf{Measure} & \textbf{Human mean (SD)} & \textbf{Default LLM mean (SD)} & \textbf{Raw gap} & \textbf{Standardized gap} \\\hline\up
Unique categories & $26.34\;(1.08)$ & $18.76\;(1.15)$ & $7.58$ & $6.82$ \\
Unique category combinations & $59.22\;(4.52)$ & $39.54\;(3.18)$ & $19.68$ & $5.03$ \\
Mean pairwise category distance & $2.403\;(0.051)$ & $2.050\;(0.056)$ & $0.353$ & $6.62$ \\
Method~1: unique concepts (\%) & $99.32\;(1.72)$ & $86.39\;(3.59)$ & $12.93$ pp & $4.59$ \\
Method~2: collective diversity & $11.53\;(0.37)$ & $7.01\;(0.12)$ & $4.52$ & $16.40$ \\\hline
\end{tabular}}
\parbox{0.9\textwidth}{\RaggedRight\vspace{6pt}\noindent {\noindent\textit{Note.} Each draw selects $10$ of $99$ participants without replacement within each condition and retains all $10$ ideas per participant. The raw gap is the human mean minus the default GPT-4o mean. The standardized gap divides the raw gap by the pooled SD of the two pool-score distributions, $\sqrt{(s_{\mathrm{Human}}^2+s_{\mathrm{LLM}}^2)/2}$. Draws overlap, so these values are descriptive and should not be interpreted using individual-level effect-size benchmarks. Method~1 uses the Universal Sentence Encoder with $\theta=.80$; Method~2 uses \texttt{text-embedding-ada-002}. For Method~1, scores and SDs are percentages and the raw gap is in percentage points (pp).}}
\end{table}

\clearpage
\primaryheading{Appendix \appCreativePersonas{}: Creative Personas Dataset} 
The following 99 personas of creative entrepreneurs in the fitness product space were systematically curated by identifying innovative founders across the industry, with the assistance of ChatGPT (\texttt{gpt-5.2}):

\begin{personatable}
\begin{longtable}{@{}L{0.32in}L{1.3in}L{\dimexpr\textwidth-1.62in-4\tabcolsep\relax}@{}}
\toprule
\textbf{No.} & \textbf{Persona} & \textbf{Description} \\
\midrule
\endfirsthead
\toprule
\textbf{No.} & \textbf{Persona} & \textbf{Description} \\
\midrule
\endhead
\midrule
\multicolumn{3}{r}{\textit{Continued on next page}} \\
\endfoot
\bottomrule
\endlastfoot
1 & John Foley & Peloton co-founder, reinvented home fitness through connected bikes and live classes \\[1pt]
2 & James Park & Fitbit co-founder, pioneered consumer wearable fitness tracking \\[1pt]
3 & Will Ahmed & WHOOP founder, reframed fitness around recovery and biometric strain \\[1pt]
4 & Harpreet Rai & Oura co-founder, made sleep and recovery core fitness metrics \\[1pt]
5 & Aly Orady & Tonal founder, brought AI-powered strength training into the home \\[1pt]
6 & Brynn Putnam & Mirror founder, blended live coaching with smart displays \\[1pt]
7 & Bruce Smith & Hydrow founder, recreated elite rowing through connected hardware \\[1pt]
8 & Eric Min & Zwift founder, turned indoor training into multiplayer virtual worlds \\[1pt]
9 & Payal Kadakia & ClassPass founder, unlocked flexible access to boutique fitness \\[1pt]
10 & Ben Francis & Gymshark founder, built a digital-native fitness apparel empire \\[1pt]
11 & Chip Wilson & Lululemon founder, transformed yoga apparel into a lifestyle brand \\[1pt]
12 & Kevin Plank & Under Armour founder, reinvented performance athletic apparel \\[1pt]
13 & Phil Knight & Nike co-founder, fused performance innovation with cultural branding \\[1pt]
14 & Jean-Luc Diard & HOKA co-founder, popularized maximalist running shoes \\[1pt]
15 & Caspar Coppetti & On Running co-founder, reimagined running shoes with cloud cushioning \\[1pt]
16 & Greg Glassman & CrossFit founder, built a global functional fitness movement \\[1pt]
17 & Elizabeth Cutler & SoulCycle co-founder, created community-driven indoor cycling \\[1pt]
18 & Julie Rice & SoulCycle co-founder, built the boutique fitness studio model into a cultural phenomenon \\[1pt]
19 & Aubrey Marcus & Onnit founder, built a ``Total Human Optimization'' fitness brand spanning supplements, functional foods, and training equipment \\[1pt]
20 & Ellen Latham & Orangetheory founder, popularized heart-rate-based interval training \\[1pt]
21 & Chris Froggatt & F45 Training founder, franchised functional group workouts globally \\[1pt]
22 & Chuck Runyon & Anytime Fitness co-founder, scaled 24/7 gym access worldwide \\[1pt]
23 & Michael Grondahl & Planet Fitness co-founder, democratized gym access with low-cost memberships \\[1pt]
24 & Mark Mastrov & 24 Hour Fitness founder, scaled big-box gyms globally \\[1pt]
25 & Jeff Zwiefel & Life Time Fitness CEO, merged health clubs with lifestyle wellness \\[1pt]
26 & Randy Hetrick & TRX founder, invented suspension training from military origins \\[1pt]
27 & Mike Cardenas & Rogue Fitness founder, scaled functional training equipment \\[1pt]
28 & Anthony Katz & Hyperice founder, commercialized recovery technology \\[1pt]
29 & Benjamin Nazarian & Therabody founder, mainstreamed percussive massage devices \\[1pt]
30 & Carl Daikeler & Beachbody founder, pioneered subscription-based home workouts \\[1pt]
31 & Tony Horton & P90X creator, revolutionized home fitness with structured video workout programs \\[1pt]
32 & Mehmet Yilmaz & Freeletics co-founder, delivered AI-driven bodyweight training to millions \\[1pt]
33 & Saeju Jeong & Noom co-founder, applied behavioral science to weight management \\[1pt]
34 & Matteo Franceschetti & Eight Sleep co-founder, optimized sleep environments for performance through temperature-controlled smart mattresses \\[1pt]
35 & Ian McCaig & Sweat co-founder, scaled mobile-first women's fitness \\[1pt]
36 & Tobi Pearce & Sweat co-founder, built a global women's fitness app empire alongside Kayla Itsines \\[1pt]
37 & Mike Lee & MyFitnessPal co-founder, scaled nutrition and calorie tracking \\[1pt]
38 & Mark Gainey & Strava co-founder, gamified endurance sports socially \\[1pt]
39 & Jason Jacobs & Runkeeper founder, enabled early GPS-based running tracking \\[1pt]
40 & Robin Thurston & MapMyFitness founder, aggregated multi-sport fitness data \\[1pt]
41 & Cliff Pemble & Garmin CEO, integrated GPS and health analytics into wearables \\[1pt]
42 & Andy Puddicombe & Headspace co-founder, integrated mindfulness into recovery \\[1pt]
43 & Michael Acton Smith & Calm co-founder, scaled mental wellness alongside fitness \\[1pt]
44 & Rick Stollmeyer & Mindbody co-founder, built software infrastructure for fitness studios \\[1pt]
45 & Alexey Lysenko & Jefit founder, built strength training tracking communities \\[1pt]
46 & Luke Zocchi & Centr co-founder, unified training, nutrition, and mindfulness \\[1pt]
47 & Adam Goldenberg & Fabletics co-founder, built data-driven activewear brands \\[1pt]
48 & Ty Haney & Outdoor Voices founder, reframed fitness as everyday movement \\[1pt]
49 & Joe De Sena & Spartan Race founder, built mass-participation endurance fitness \\[1pt]
50 & Samir Goel & Fittr founder, scaled online coaching for strength training \\[1pt]
51 & Gautam Thakur & HealthifyMe co-founder, combined AI nutrition with fitness tracking \\[1pt]
52 & Ralf Wenzel & Gymondo co-founder, expanded digital fitness platforms in Europe \\[1pt]
53 & Moawia Eldeeb & Les Mills digital leader, transitioned group fitness to streaming \\[1pt]
54 & Dave Jenkins & Obé Fitness co-founder, built personality-driven digital workouts \\[1pt]
55 & Ethan Eismann & Aaptiv co-founder, pioneered audio-based fitness coaching \\[1pt]
56 & Derek Flanzraich & Greatist founder, blended fitness, health, and content \\[1pt]
57 & Nick Woodman & GoPro founder, enabled action-sport fitness content creation \\[1pt]
58 & James Murdoch & Calm investor-chair, scaled wellness-tech ecosystems \\[1pt]
59 & Brian Bordainick & SweatPals founder, built social fitness discovery platforms \\[1pt]
60 & Rory McIlroy & TMRW Sports co-founder, blended elite training with tech-enabled venues \\[1pt]
61 & David Giampaolo & BODi leader, evolved Beachbody into an interactive fitness platform \\[1pt]
62 & Shervin Pishevar & Early fitness-tech investor, accelerated connected fitness startups \\[1pt]
63 & Nick Caldwell & Looker product leader, influenced fitness analytics ecosystems \\[1pt]
64 & Emily Heyward & Brand strategist shaping modern fitness product identities \\[1pt]
65 & Jeff Halevy & EXOS executive, integrated elite performance science into fitness \\[1pt]
66 & Nerio Alessandri & Technogym founder, merged luxury fitness equipment with digital ecosystems \\[1pt]
67 & Don Faul & Athletic Brewing CEO, connected fitness culture with lifestyle brands \\[1pt]
68 & Chris Hemsworth & Centr co-founder, blended celebrity influence with digital fitness \\[1pt]
69 & Kayla Itsines & Sweat creator, scaled global fitness programs via apps \\[1pt]
70 & Justin Kan & Health and fitness tech investor shaping new platforms \\[1pt]
71 & Ryan Holmes & Recovery-focused fitness investor and entrepreneur \\[1pt]
72 & Todd Sinett & Posture-focused fitness entrepreneur integrating mobility \\[1pt]
73 & Jeff Galloway & Run-walk training entrepreneur, scaled endurance fitness programs \\[1pt]
74 & Sam Parr & Fitness media entrepreneur building community-driven health brands \\[1pt]
75 & Natalie Jill & Digital fitness entrepreneur targeting midlife wellness \\[1pt]
76 & Sebastian Thrun & Health-tech entrepreneur applying AI to fitness and longevity \\[1pt]
77 & Max Levchin & Fitness-tech investor backing connected health platforms \\[1pt]
78 & Daniel Ek & Spotify founder influencing fitness through audio training ecosystems \\[1pt]
79 & Kevin Rose & Early Quantified Self entrepreneur shaping fitness data culture \\[1pt]
80 & Esther Dyson & Wellness-tech investor advancing preventive fitness models \\[1pt]
81 & Naval Ravikant & Angel investor backing fitness and longevity startups \\[1pt]
82 & Shan-Lyn Ma & Zola co-founder, applied habit-building design principles relevant to wellness platforms \\[1pt]
83 & Yoni Assia & eToro founder, backed social fitness investing platforms \\[1pt]
84 & Hank Haney & Golf fitness training entrepreneur scaling performance coaching \\[1pt]
85 & Chris Paul & CP3 performance brand co-founder, merged elite training with consumer fitness \\[1pt]
86 & David Barton & David Barton Gym founder, blended nightlife aesthetics with gym experiences \\[1pt]
87 & Steven Plofker & Blink Fitness co-founder, democratized affordable boutique-style gyms \\[1pt]
88 & Danny Harris & Alo Yoga co-founder, built a premium yoga/athleisure brand that scaled from apparel into a broader wellness ecosystem \\[1pt]
89 & Anne Mahlum & [solidcore] founder, built a high-intensity, low-impact Pilates-inspired boutique studio chain and scaled it nationally \\[1pt]
90 & Barry Jay & Barry's co-founder, launched the original Barry's Bootcamp concept and helped spark the modern boutique HIIT studio movement \\[1pt]
91 & Rob Deutsch & Fhitting Room co-founder, scaled strength-based boutique gyms \\[1pt]
92 & Noam Tamir & TS Fitness founder, integrated personal training with digital tracking \\[1pt]
93 & Jay Cardiello & Cardiello Concepts founder, branded functional training systems \\[1pt]
94 & Cesar Carvalho & Wellhub (Gympass) co-founder, connected employees to global fitness access \\[1pt]
95 & Katrina Scott & Tone It Up co-founder, built women-focused digital fitness programs \\[1pt]
96 & Khalil Zahar & FightCamp founder and CEO, turned boxing training into connected home fitness with punch-tracking technology \\[1pt]
97 & Adam Peake & FitXR co-founder, merged VR gaming with immersive fitness workouts \\[1pt]
98 & Michael Nash & Rumble Boxing co-founder, scaled music-driven boxing fitness studios \\[1pt]
99 & Mike Doehla & Stronger U founder, built data-driven nutrition and coaching platform \\[1pt]
\end{longtable}\addtocounter{table}{-1}
\end{personatable}

\clearpage
\primaryheading{Appendix \appOrdinaryPersonas{}: Ordinary Personas Dataset}

The following 99 personas were randomly selected from \citet{ge2024scaling} to represent diverse social roles, professions, and everyday contexts.

\begin{personatable}
\begin{longtable}{@{}L{0.32in}L{\dimexpr\textwidth-0.32in-2\tabcolsep\relax}@{}}
\toprule
\textbf{No.} & \textbf{Persona description} \\
\midrule
\endfirsthead
\toprule
\textbf{No.} & \textbf{Persona description} \\
\midrule
\endhead
\midrule
\multicolumn{2}{r}{\textit{Continued on next page}} \\
\endfoot
\bottomrule
\endlastfoot
1 & A retired politician who has successfully campaigned for library funding in the past \\[1pt]
2 & A newbie software engineer not familiar with Sphinx search server \\[1pt]
3 & A Python developer with a keen interest in simplifying complex array operations using NumPy for high-performance computing \\[1pt]
4 & A die-hard football fan and avid follower of South American football clubs, with a special interest in scrapbooking historical club information \\[1pt]
5 & A museum curator who provides access to rare and valuable documents detailing the social dynamics of European court life \\[1pt]
6 & A government official responsible for supporting research on language contact and bilingualism \\[1pt]
7 & A hospital CEO seeking to build a positive public image and manage crisis situations with the help of the executive \\[1pt]
8 & A small business owner in Memphis, TN who is skeptical of IT services \\[1pt]
9 & A supportive father who appreciates the discipline and artistry of ballet \\[1pt]
10 & A veterinary technician who provides advice and support for animal health concerns in farming \\[1pt]
11 & A screenwriter who has written episodes for some of the most iconic sitcoms in television history \\[1pt]
12 & A music theory student who helps the younger sibling understand complex guitar chords and scales \\[1pt]
13 & A traditional music ensemble leader dedicated to preserving heritage through melody \\[1pt]
14 & A project manager who relies on economic analysis to guide investment decisions \\[1pt]
15 & A quantum physicist who believes that traditional biology fails to explain complex life processes \\[1pt]
16 & A knowledgeable resident of each community who provides insider information and cultural context \\[1pt]
17 & A full-stack developer who is passionate about creating robust data visualization tools \\[1pt]
18 & An ethics philosopher who challenges the retired judge's perspectives and helps broaden their understanding \\[1pt]
19 & An eccentric street-art equipment vendor who knows the ins-and-outs of the local graffiti scene \\[1pt]
20 & An international student who is interested in attaining scholarships and studying in Arab regions \\[1pt]
21 & A recent divorcee looking for inspiration and practical advice on starting a new business \\[1pt]
22 & A paleontologist who studies ancient mammal fossils to understand their feeding habits \\[1pt]
23 & A hardworking manual laborer who inspires their children with their perseverance and tenacity \\[1pt]
24 & A representative from a government agency funding research on social network analysis for policy-making purposes \\[1pt]
25 & A retired pyrotechnician with a fascination for fireworks and explosives \\[1pt]
26 & An anthropologist conducting fieldwork on cultural syncretism in Eastern Europe \\[1pt]
27 & A military history blogger who provides valuable insights and research on the historical accuracy of military model kits \\[1pt]
28 & A business strategist with extensive knowledge of risk management and corporate governance, providing insights on cybersecurity measures \\[1pt]
29 & A casting director who helps connect talent scouts with aspiring actors seeking dialect training \\[1pt]
30 & A retired accountant who willingly participates in experiments and interviews to contribute to the study \\[1pt]
31 & A data manager responsible for ensuring the accuracy and integrity of clinical trial databases \\[1pt]
32 & An aspiring young golfer from a rural town, deeply inspired by underdog stories and achievement against the odds \\[1pt]
33 & An experienced academic advisor who worked with Chris Shelton during his MSc program \\[1pt]
34 & A fellow lab technician who loves creating color-coded labeling systems and organizing lab supplies \\[1pt]
35 & A dairy farmer who occasionally collaborates with the sheep farmer to host farm tours for tourists \\[1pt]
36 & A close friend who has known Jacque Snow since the first grade \\[1pt]
37 & A computer scientist specializing in machine learning who assesses the technical aspects of the submitted papers \\[1pt]
38 & A retired politician who guides the council member on effective strategies to engage and mobilize young voters \\[1pt]
39 & A young child who embraces simplicity by finding joy in the smallest things \\[1pt]
40 & A pharmaceutical company representative interested in commercializing optimized medication formulations \\[1pt]
41 & An old lady who is very active on social media and loves gossiping about politicians \\[1pt]
42 & A charismatic and enthusiastic voice that introduces and hypes up wrestling matches \\[1pt]
43 & A visionary entrepreneur who is also in the retro-themed business but targets a different niche \\[1pt]
44 & A retired childcare professional who always has a soothing lullaby up her sleeve \\[1pt]
45 & A music history enthusiast visiting New Orleans to learn about the city's jazz roots, seeking recommendations from the local resident \\[1pt]
46 & A bride-to-be searching for the perfect traditional Indian jewelry to complete her wedding ensemble \\[1pt]
47 & A passionate historical fiction writer who is infatuated with the English countryside and its old-world charm \\[1pt]
48 & A British historian focusing on twentieth-century British civil servants \\[1pt]
49 & An individual who feels their constitutional rights were violated and seeks justice through legal representation \\[1pt]
50 & An elderly retired professor who loves to learn and is interested in understanding the concept of remote work \\[1pt]
51 & A soccer coach whose team struggles with maintaining focus and motivation during matches \\[1pt]
52 & A government official working on labor market reforms and income redistribution \\[1pt]
53 & A software developer creating apps to make historical content more accessible and interactive \\[1pt]
54 & A fellow researcher from a different university, collaborating on decoding encrypted messages \\[1pt]
55 & A local community leader who provides insights into the social and cultural dynamics of the psychologist's new environment \\[1pt]
56 & A talented guitarist eager to understand the theoretical principles behind different musical styles \\[1pt]
57 & An individual who has a deep love for free and open-source software (FOSS) \\[1pt]
58 & A naturalized U.S. citizen who respects the U.S. Constitution and takes pride in American democracy \\[1pt]
59 & A reader who frequently questions the bias and objectivity of the news articles \\[1pt]
60 & An online gaming competitor known for their skills in optimizing device performance for gaming \\[1pt]
61 & A college student who enjoys Zumba as a fun and energetic way to stay fit \\[1pt]
62 & A wealthy businesswoman who fell prey to the con artist's scheme and lost her life savings \\[1pt]
63 & A seasoned hospice nurse supervisor who provides guidance and support in managing the emotional challenges of the job \\[1pt]
64 & An aerospace engineer originally from Oklahoma now working in California \\[1pt]
65 & A person who sees bias in the media and believes it plays a role in escalating racial tensions \\[1pt]
66 & An aunt who is a music composer and helps create original flamenco compositions \\[1pt]
67 & A lifelong fan of Tanya Chua who always eagerly awaits her new releases \\[1pt]
68 & A seasoned agent who is hesitant to fully embrace the digital publishing revolution \\[1pt]
69 & A principal committed to improving college acceptance rates at their school \\[1pt]
70 & A stand-up comedian whose comedy routines are filled with profanity and controversial topics \\[1pt]
71 & An award-winning ink chemist who has revolutionized the industry with their inventions \\[1pt]
72 & A visionary car designer known for creating breakthrough electric vehicle concepts \\[1pt]
73 & An English literature professor at Lawrence University, Appleton, who is passionate about local literary history \\[1pt]
74 & A socially conscious entrepreneur who runs a vegan shoe company and shares the same vision \\[1pt]
75 & An IT director responsible for implementing and overseeing version control policies across multiple teams \\[1pt]
76 & A computer science professor skeptical about AI detection tools \\[1pt]
77 & A person with multiple sclerosis who shares their journey and coping strategies with others in a virtual community \\[1pt]
78 & A fashion vlogger who criticizes DIY methods and promotes professional haute couture \\[1pt]
79 & A fellow veteran who was inspired by Galen Cole's bravery and selflessness in combat \\[1pt]
80 & A mainstream comic book writer criticized for their lack of diversity in storylines \\[1pt]
81 & A teenager who survived a car crash and is now determined to raise awareness about the dangers of reckless driving \\[1pt]
82 & A young projectionist who shares the owner's love for vintage cinema and maintains the film projectors \\[1pt]
83 & A technology support specialist offering affordable data backup services for small businesses \\[1pt]
84 & ``Sipho,'' a 25-year-old South African political science graduate student who hopes to run for local office someday \\[1pt]
85 & An old college bandmate of Mary, now working in real estate, but still a fan of avant-garde music \\[1pt]
86 & A Conservative Manx and retired civil servant residing in the Isle of Man who is curious and enthusiastic about learning different cultures and comparing national statistics \\[1pt]
87 & A beauty blogger who is skeptical about celebrity doctors and their branded products \\[1pt]
88 & A younger sibling who is studying French and interested in bilingualism in Canada \\[1pt]
89 & An art therapist recognizing the potential of VR in creating therapeutic environments \\[1pt]
90 & A hobbyist radio DJ with their own small-scale broadcasting setup \\[1pt]
91 & A medical researcher who recognizes the importance of accurate and accessible medical records in advancing scientific knowledge \\[1pt]
92 & A veteran defender who has faced the striker numerous times and considers them a formidable opponent \\[1pt]
93 & A computer scientist specializing in artificial intelligence for information retrieval and recommendation systems \\[1pt]
94 & A backpacker from a different country who recently explored Antwerp and can offer advice on hidden gems \\[1pt]
95 & A maulana who offers guidance on integrating physical fitness with Islamic principles \\[1pt]
96 & A high school student determined to follow in their sibling's footsteps and get into Harvard \\[1pt]
97 & A competitive sports agent focused on poaching promising athletes \\[1pt]
98 & A cryptographer who owns a gaming blog and often engages in late-night gaming tournaments \\[1pt]
99 & A veteran hip hop producer who has worked with some of the biggest names in the industry \\[1pt]
\end{longtable}\addtocounter{table}{-1}
\end{personatable}

\clearpage
\primaryheading{Appendix \appPromptingStructure{}: Prompting Structure for Sequential Idea Generation with Personas}

In Study~1, each simulated participant in the LLM Creative Personas and LLM Ordinary Personas conditions received a persona in the system message and generated ten ideas sequentially in one session. The prompting procedure follows.

\begin{tcolorbox}[
    enhanced,
    colback=gray!5!white,
    colframe=gray!50!black,
    title=Prompting Structure for Sequential Idea Generation with Personas,
    listing only,
    listing engine=listings,
    breakable,
    listing options={
        basicstyle=\ttfamily\scriptsize,
        columns=fullflexible,
        keepspaces=true,
        breaklines=true,
        breakatwhitespace=true
    }
]
\begin{lstlisting}
System Message:
system_message = f"You are acting as this persona: {profile}. Generate ideas from this persona's perspective."
messages = [{"role": "system", "content": system_message}]

Sequential Idea Loop:
local_results = []
for i in range(NUM_IDEAS):
    idea_prompt = f"Give me 1 new idea for a fitness product. The idea should be explained in exactly one sentence. Just give me the idea, labeled 'Idea #{i+1}'."
    ...
\end{lstlisting}
\end{tcolorbox}

\clearpage
\primaryheading{Appendix \appSeqCot{}: CoT with Sequential Idea Generation}

\tertiaryheading{Experimental Setup}
Study~2 generated all ten ideas at once, following \citet{defreitas2025ideation}. This robustness check instead applies CoT revision to sequentially generated Study~1 ideas. Using the main-text generator, GPT-4o (\texttt{gpt-4o-2024-11-20}), we compare \textit{LLM Default $+$ CoT (Sequential)} and \textit{LLM Ordinary Personas $+$ CoT (Sequential)} with their corresponding main-study baselines.

\vspace{.5em}
\tertiaryheading{LLM Conditions} We applied CoT revisions to two different sources of base ideas.

\noindent (1) LLM CoT Revision. We entered each Study~1 LLM participant's ($N=99$) ten original, sequentially generated ideas directly into the CoT revision prompt.

\begin{tcolorbox}[
    enhanced,
    colback=gray!5!white,
    colframe=gray!50!black,
    title=Default $\rightarrow$ CoT Revision Prompt (inputs the 10 sequential ideas from Study 1: LLM Default),
    listing only,
    listing engine=listings,
    breakable,
    listing options={
        basicstyle=\ttfamily\scriptsize,
        columns=fullflexible,
        keepspaces=true,
        breaklines=true,
        breakatwhitespace=true
    }
]
\begin{lstlisting}
Please review the 10 fitness product ideas below. Revise them to make them bolder and more different from one another. Ensure that no two ideas are the same across your ten responses.

Idea #1: ...
Idea #2: ...
...
Idea #10: ...

Return ONLY a JSON array with 10 objects, each with "idea_number" (1-10) and "idea_content" (revised one-sentence description). Make them bolder and more distinctive.
\end{lstlisting}
\end{tcolorbox}

\noindent (2) LLM Personas CoT Revision. We applied the same revision procedure to the ten original, sequentially generated ideas of each Study~1 persona participant ($N=99$), reminding the model of its assigned persona.

\begin{tcolorbox}[
    enhanced,
    colback=gray!5!white,
    colframe=gray!50!black,
    title=Personas $\rightarrow$ CoT Revision Prompt (inputs the 10 sequential ideas from Study 1: LLM Personas),
    listing only,
    listing engine=listings,
    breakable,
    listing options={
        basicstyle=\ttfamily\scriptsize,
        columns=fullflexible,
        keepspaces=true,
        breaklines=true,
        breakatwhitespace=true
    }
]
\begin{lstlisting}
You are acting as this persona: {persona}.

Please review the 10 fitness product ideas below that you previously generated from this persona's perspective. Revise them to make them bolder and more different from one another while staying faithful to the persona. Ensure that no two ideas are the same across your ten responses.

Idea #1: ...
Idea #2: ...
...
Idea #10: ...

Return ONLY a JSON array with 10 objects, each with "idea_number" (1-10) and "idea_content" (revised one-sentence description). Make them bolder and more distinctive.
\end{lstlisting}
\end{tcolorbox}

\noindent Both conditions required \emph{bolder}, \emph{more distinct} revisions and JSON-only output.

\vspace{1em}
\tertiaryheading{Idea Diversity}
Table~\ref{tab:sequential_full_sample_diversity} uses the first 83 participants per condition (830 ideas), matching the main-text subsample. Sequential CoT raises default-LLM unique combinations from 85 to 103, with categories unchanged at 23; humans reach 196 combinations. Personas alone yield 27 categories and 190 combinations, versus humans' 28 and 196. Personas $+$ sequential CoT yield 27 categories and 203 combinations, the strongest LLM result and slightly above humans on combinations. Study~1 baselines use main-text values; sequential-CoT values come from a separate categorization run and may differ trivially across runs (Appendix~\appModelChoice{}).

\begin{table}[!htbp]
\centering\normalsize\singlespacing\setlength{\tabcolsep}{4pt}\renewcommand{\arraystretch}{1.2}\setlength{\RaggedRightParindent}{0pt}\setlength{\parindent}{0pt}
\caption{\textbf{Full Sample Comparison: Diversity Across Human and LLM Conditions (Sequential Generation $+$ CoT Revision).} Each condition uses the first 83 participants (830 ideas), matching the main-text subsample.}\label{tab:sequential_full_sample_diversity}
\begin{tabularx}{\textwidth}{@{}>{\RaggedRight\arraybackslash}X>{\centering\arraybackslash}p{1.20in}>{\centering\arraybackslash}p{1.20in}>{\centering\arraybackslash}p{0.80in}@{}}
\hline\up
\textbf{Group} & \textbf{Total Categories} & \textbf{Unique Combinations} & \textbf{Total Ideas} \\\hline\up
Human (Study 1) & 28 & 196 & 830 \\
LLM Default (Study 1) & 23 & 85 & 830 \\
LLM Default $+$ CoT (Sequential) & 23 & 103 & 830 \\
LLM Ordinary Personas (Study 1) & 27 & 190 & 830 \\
LLM Ordinary Personas $+$ CoT (Sequential) & 27 & 203 & 830 \down\\\hline
\end{tabularx}

\end{table}

\begin{samepage}
Bootstrap comparisons (Tables~\ref{tab:sequential-default-cot-comparison} and~\ref{tab:sequential-persona-cot-comparison}) show:
\begin{itemize}
\item Default $\rightarrow$ CoT: categories $T_{\text{cat}}=18.76$ ($SE=0.11$) $\rightarrow 19.36$ ($SE=0.11$), $p<.001$; combinations $T_{\text{comb}}=39.54$ ($SE=0.32$) $\rightarrow 41.24$ ($SE=0.27$), $p<.001$; distance $d=2.05$ ($SE=0.01$) $\rightarrow 2.13$ ($SE=0.00$), $p<.001$.
\item Ordinary personas $\rightarrow$ ordinary personas $+$ CoT: categories $T_{\text{cat}}=24.35$ ($SE=0.13$) $\rightarrow 25.04$ ($SE=0.12$), $p<.001$; combinations $T_{\text{comb}}=57.37$ ($SE=0.58$) $\rightarrow 57.64$ ($SE=0.54$), $p=.734$; distance $d=2.36$ ($SE=0.01$) $\rightarrow 2.35$ ($SE=0.01$), $p=.388$. Only categories increase significantly.
\end{itemize}
\end{samepage}

\begin{table}[!htbp]
\centering\normalsize\singlespacing\setlength{\tabcolsep}{4pt}\renewcommand{\arraystretch}{1.2}\setlength{\RaggedRightParindent}{0pt}\setlength{\parindent}{0pt}
\caption{\textbf{Bootstrapped Diversity Comparison: LLM Default vs.\ LLM Default $+$ CoT (Mean $\pm$ SE).} Means and $SE$s over $100$ bootstrap draws of $10$ participants $\times$ $10$ ideas; $p$-values are Welch $t$-tests on the bootstrap distributions.}\label{tab:sequential-default-cot-comparison}
\begin{tabularx}{\textwidth}{@{}>{\RaggedRight\arraybackslash}X>{\centering\arraybackslash}p{1.20in}>{\centering\arraybackslash}p{1.20in}>{\centering\arraybackslash}p{0.80in}@{}}
\hline\up
\textbf{Metric} & \textbf{LLM Default (Study 1)} & \textbf{LLM Default $+$ CoT (Sequential)} & \textbf{$p$-value} \\\hline\up
Total Unique Categories (10 $\times$ 10 Ideas) & $18.76 \pm 0.11$ & $19.36 \pm 0.11$ & $< .001$ \\
Total Unique Combinations (10 $\times$ 10 Ideas) & $39.54 \pm 0.32$ & $41.24 \pm 0.27$ & $< .001$ \\
Average Pairwise Distance (10 $\times$ 10 Ideas) & $2.05 \pm 0.01$ & $2.13 \pm 0.00$ & $< .001$ \down\\\hline
\end{tabularx}

\end{table}

\begin{table}[!htbp]
\centering\normalsize\singlespacing\setlength{\tabcolsep}{4pt}\renewcommand{\arraystretch}{1.2}\setlength{\RaggedRightParindent}{0pt}\setlength{\parindent}{0pt}
\caption{\textbf{Bootstrapped Diversity Comparison: LLM Ordinary Personas vs.\ LLM Ordinary Personas $+$ CoT (Mean $\pm$ SE).} Means and $SE$s over $100$ bootstrap draws of $10$ participants $\times$ $10$ ideas; $p$-values are Welch $t$-tests on the bootstrap distributions. Baseline (Study 1) values are the main-text values; sequential-CoT values come from the sequential-CoT analysis run.}\label{tab:sequential-persona-cot-comparison}
\begin{tabularx}{\textwidth}{@{}>{\RaggedRight\arraybackslash}X>{\centering\arraybackslash}p{1.20in}>{\centering\arraybackslash}p{1.20in}>{\centering\arraybackslash}p{0.80in}@{}}
\hline\up
\textbf{Metric} & \textbf{LLM Ordinary Personas (Study 1)} & \textbf{LLM Ordinary Personas $+$ CoT (Sequential)} & \textbf{$p$-value} \\\hline\up
Total Unique Categories (10 $\times$ 10 Ideas) & $24.35 \pm 0.13$ & $25.04 \pm 0.12$ & $< .001$ \\
Total Unique Combinations (10 $\times$ 10 Ideas) & $57.37 \pm 0.58$ & $57.64 \pm 0.54$ & $.734$ \\
Average Pairwise Distance (10 $\times$ 10 Ideas) & $2.36 \pm 0.01$ & $2.35 \pm 0.01$ & $.388$ \down\\\hline
\end{tabularx}

\end{table}

\begin{figure}[!htbp]
\caption{FIXATION ANALYSIS: DIVERSITY ACCUMULATION FOR LLMS WITH AND WITHOUT CHAIN-OF-THOUGHT AND PERSONAS (SEQUENTIAL GENERATION)}\label{fig:sequential_llm_fixation_seq}
\centering
\includegraphics[width=\textwidth]{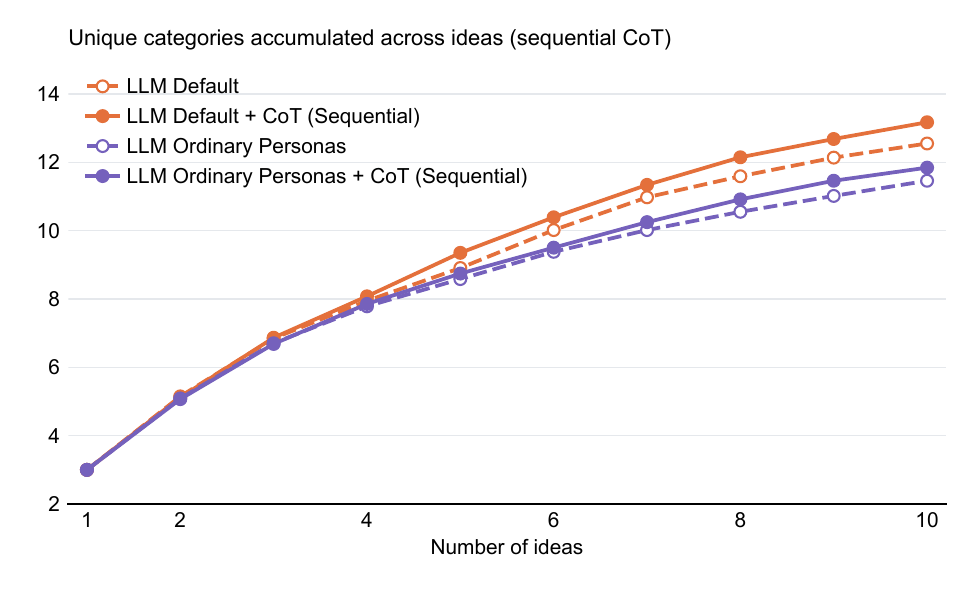}
\fignote{$\beta$ is the average slope from linear regressions on each participant's category accumulation curve; comparisons are Welch $t$-tests on participant-level slopes: $\beta_{\text{default}} = 1.023$ vs.\ $\beta_{\text{CoT}} = 1.105$ ($p = .001$); $\beta_{\text{ordinary}} = 0.875$ vs.\ $\beta_{\text{ordinary CoT}} = 0.930$ ($p = .178$).}

\end{figure}

\vspace{1em}
\tertiaryheading{Fixation Analysis}
CoT reduces category fixation for the default LLM (Figure~\ref{fig:sequential_llm_fixation_seq}): slopes rise from $\beta_{\text{default}}=1.023$ ($SE=0.017$) to $\beta_{\text{CoT}}=1.105$ ($SE=0.018$), $p=.001$. With personas, the increase is not significant: $\beta_{\text{ordinary}}=0.875$ ($SE=0.028$) versus $\beta_{\text{ordinary+CoT}}=0.930$ ($SE=0.029$), $p= .178$.

\clearpage
\primaryheading{Appendix \appCotPrompts{}: Chain-of-Thought Prompts (Study 2)}

The original Chain-of-Thought prompt from \citet{defreitas2025ideation} and the exact Study~2 prompts for LLM CoT and LLM Ordinary Personas CoT follow.

\begin{tcolorbox}[
    enhanced,
    colback=gray!5!white,
    colframe=gray!50!black,
    title=CoT Prompt from De Freitas et al.\ (2025),
    listing only,
    listing engine=listings,
    breakable,
    listing options={
        basicstyle=\ttfamily\scriptsize,
        columns=fullflexible,
        keepspaces=true,
        breaklines=true,
        breakatwhitespace=true
    }
]
\begin{lstlisting}
Follow these steps. Do each step, even if you think you do not need to.
First, generate a list of 30 ideas (short title only).
Second, go through the list and determine whether the ideas are different and bold. Modify the ideas as needed to make them bolder and more different. No two ideas should be the same. This is important!
Next, give the ideas a name and combine each with a paper description. The name and idea are separated by a colon and followed by a description. The idea should be expressed as a paragraph of 40--80 words.
Do this step by step!
\end{lstlisting}
\end{tcolorbox}

\begin{tcolorbox}[
    enhanced,
    colback=gray!5!white,
    colframe=gray!50!black,
    title=LLM CoT Prompt,
    listing only,
    listing engine=listings,
    breakable,
    listing options={
        basicstyle=\ttfamily\scriptsize,
        columns=fullflexible,
        keepspaces=true,
        breaklines=true,
        breakatwhitespace=true
    }
]
\begin{lstlisting}
Generate fitness product ideas. Follow these steps:
1. Generate a list of 10 ideas (short title only)
2. Review and make them bolder and more different - no two ideas should be the same
3. Return ONLY a JSON array with 10 objects, each with "idea_number" (1-10) and "idea_content" (one sentence description)

Return ONLY the JSON array, no other text.
\end{lstlisting}
\end{tcolorbox}

\begin{tcolorbox}[
    enhanced,
    colback=gray!5!white,
    colframe=gray!50!black,
    title=LLM Ordinary Personas CoT Prompt,
    listing only,
    listing engine=listings,
    breakable,
    listing options={
        basicstyle=\ttfamily\scriptsize,
        columns=fullflexible,
        keepspaces=true,
        breaklines=true,
        breakatwhitespace=true
    }
]
\begin{lstlisting}
You are acting as this persona: {persona}

Generate fitness product ideas from this persona's perspective using chain of thought reasoning. Follow these steps:

1. First, think about what this persona would value in fitness products based on their characteristics.
2. Generate a list of 10 fitness product ideas that align with this persona's needs and preferences (short titles only).
3. Review and refine ideas to make them bolder and more different - ensure that no two ideas are the same.
4. Consider how this persona would uniquely approach or modify each idea.

Return ONLY a JSON array with 10 objects, each with:
- "idea_number" (1-10)
- "idea_content" (one sentence description from this persona's perspective)

Return ONLY the JSON array, no other text or explanation.
\end{lstlisting}
\end{tcolorbox}

Human CoT participants generated ten short summaries on one survey page, then revised each to differ as much as possible from its initial version (Figure~\ref{fig:human_cot_1}).

\begin{figure}[!tbp]
\caption{HUMAN CHAIN-OF-THOUGHT CONDITION SURVEY INTERFACE}\label{fig:human_cot_1}
\centering
\includegraphics[width=0.9\textwidth,keepaspectratio]{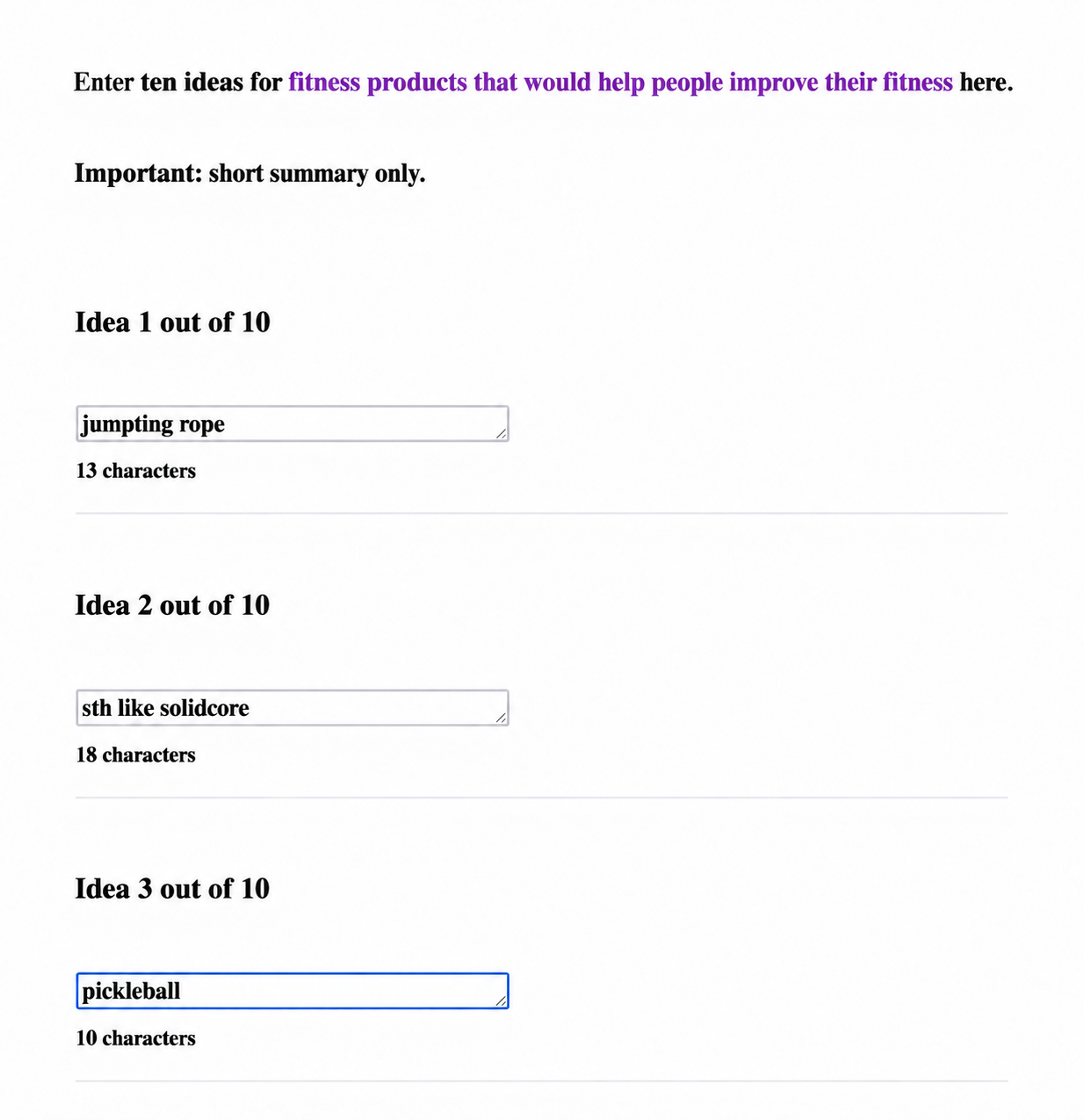}
\par\textbf{(a) Step 1: Participants generated 10 short summaries.}
\fignote{Screenshots of the two-step CoT procedure administered to human participants: (a) generate ten short idea summaries, then (b) revise the summaries to be as different as possible before expanding them into full ideas.}
\end{figure}

\begin{figure}[!tbp]
\ContinuedFloat
\caption[]{HUMAN CHAIN-OF-THOUGHT CONDITION SURVEY INTERFACE (CONTINUED)}
\centering
\includegraphics[width=0.9\textwidth,keepaspectratio]{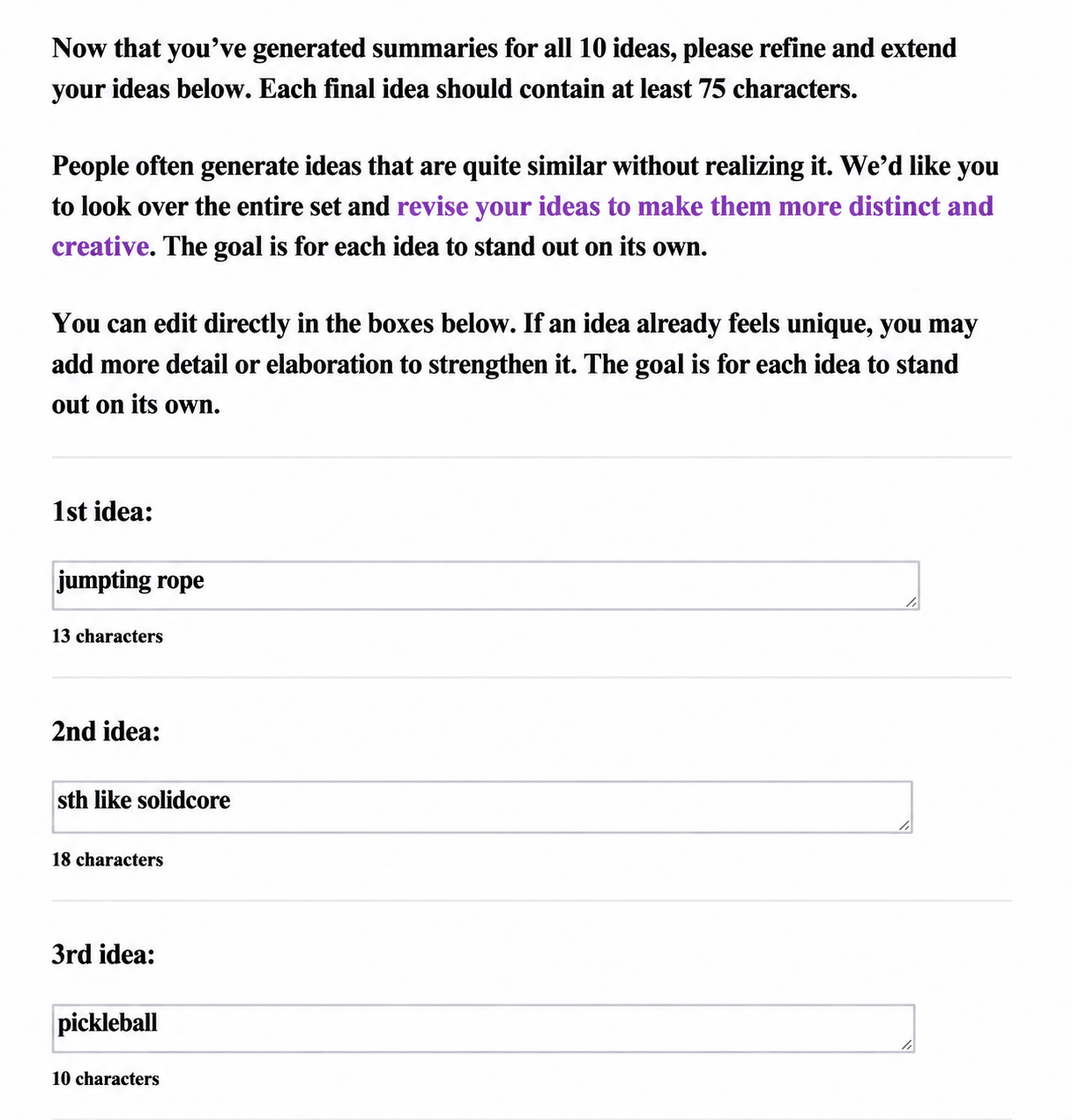}
\par\textbf{(b) Step 2: Participants revised summaries to be as different as possible.}
\end{figure}

\clearpage
\primaryheading{Appendix \appPartitioning{}: Knowledge Partitioning Analysis in Study 2}

\begin{figure}[!htbp]
\caption{STUDY 2 FIRST IDEAS ONLY FOR DIFFERENT LLM CONDITIONS}\label{fig:study4firstideas}
\centering
\includegraphics[width=\textwidth]{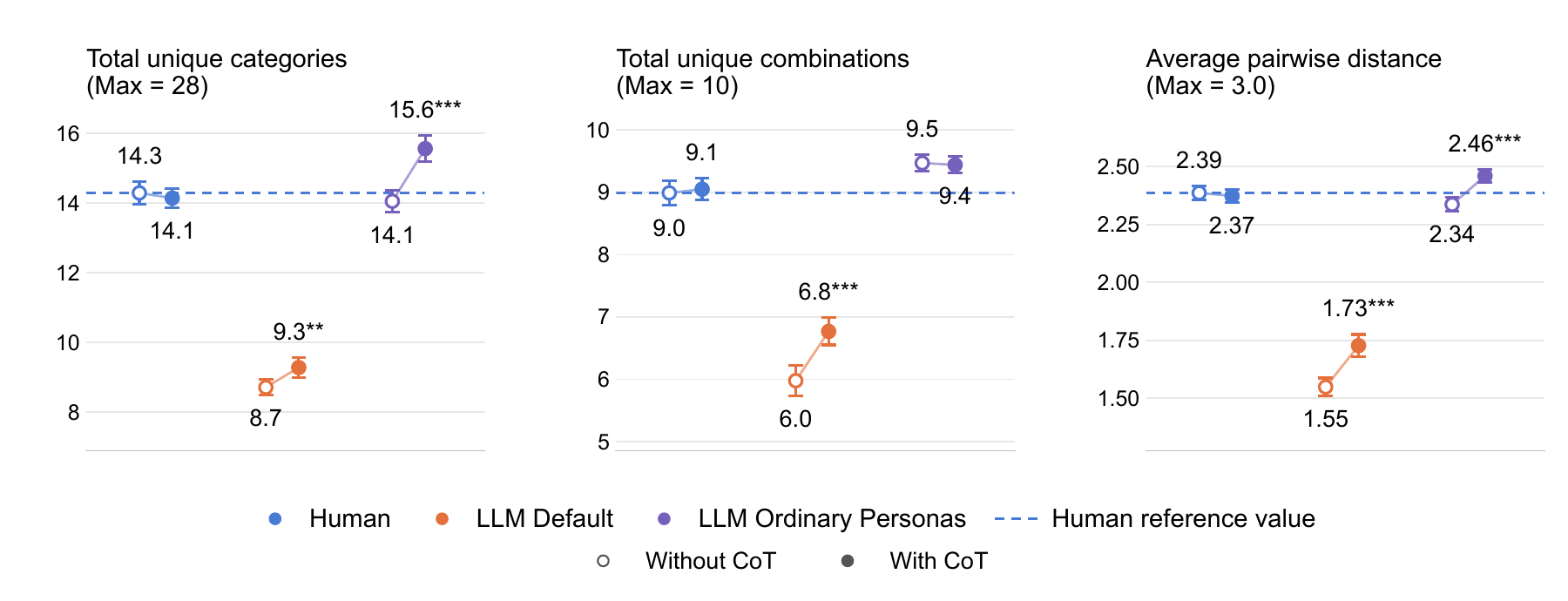}
\fignote{Means from $100$ bootstrap draws of $10$ participants' first ideas each; error bars are $95\%$ confidence intervals ($\pm 1.96\,SE$). Open markers show baseline conditions and filled markers the corresponding CoT conditions. Stars mark Welch $t$-tests on the bootstrap distributions comparing baseline vs.\ CoT within each ideator type: $^{***}p<.001$, $^{**}p<.01$, $^{*}p<.05$; unmarked comparisons are not significant. Light connecting lines link condition means and do not denote matched participants. The blue dashed line marks the Human baseline reference value. Panel titles report theoretical maxima for the sampled idea pool; vertical scales differ across metrics and align the human reference at the same height.}

\end{figure}

\tertiaryheading{Knowledge Partitioning (First Ideas)}
CoT modestly improves all first-idea metrics (Figure~\ref{fig:study4firstideas}):
\begin{itemize}
\item Categories: default $T_{\text{cat}}=8.71$ ($SE=0.11$) vs.\ CoT $9.28$ ($SE=0.15$), $p=.003$.
\item Combinations: default $T_{\text{comb}}=5.98$ ($SE=0.12$) vs.\ CoT $6.77$ ($SE=0.11$), $p<.001$.
\item Distance: default $d=1.55$ ($SE=0.02$) vs.\ CoT $1.73$ ($SE=0.02$), $p<.001$.
\end{itemize}
CoT remains well below Human and persona conditions, recovering only part of the starting-point gap.

\tertiaryheading{Knowledge Partitioning (Embedding Analysis)}
Ordinary personas have the highest between-participant variation ($M=0.635$, $SD=0.045$); CoT modestly exceeds default ($M=0.318$, $SD=0.010$ vs.\ $M=0.241$, $SD=0.014$). These results indicate that CoT recovers only part of the between-participant variation observed with personas.

\end{document}